\documentclass{revtex4}

\usepackage{graphicx}
\usepackage{amsmath,amssymb}
\usepackage{bm}
\usepackage{textcomp}

\newcommand{\bvec}{\boldsymbol}

\begin{document}

\preprint{KUNS-2505}

\title{Description of $\alpha$-cluster tail in  
$^{8}$Be and $^{20}$Ne: Delocalization of 
$\alpha$ cluster because of quantum penetration}
\author{Yoshiko Kanada-En'yo}
\affiliation{Department of Physics, Kyoto University, Kyoto 606-8502, Japan}
\begin{abstract}
We analyze the $\alpha$-cluster wave functions in cluster states of $^8$Be and $^{20}$Ne 
by comparing 
 the exact relative wave function obtained by 
the generator coordinate method (GCM) with
various types of trial functions. For the trial functions, 
we adopt the fixed range shifted Gaussian of the Brink-Bloch (BB) wave function, 
the spherical Gaussian with the adjustable range parameter 
of the spherical Thosaki-Horiuchi-Schuck-R\"opke 
(sTHSR), the deformed Gaussian of the deformed THSR (dTHSR), 
and a function with the Yukawa tail (YT).
The quality of the description of the exact wave function with a trial function
is judged by the squared overlap between the trial function and the GCM wave function. 
The better result is obtained 
with the sTHSR wave function than the BB wave function, and further improvement can be 
done with the dTHSR wave function because these wave functions can 
describe the outer tail better.
The YT wave function gives almost the equal quality to 
or even better quality than the dTHSR wave function indicating that the outer tail of $\alpha$ cluster states
is characterized by the Yukawa-like tail 
rather than the Gaussian tail.
In the weakly bound $\alpha$ cluster states with the small 
$\alpha$ separation energy and the low centrifugal and Coulomb barriers, the outer tail part is the slowly damping 
function described well by the quantum penetration through the effective barrier. 
This outer tail characterizes the 
almost zero-energy free $\alpha$ gas behavior, i.e., 
the delocalization of cluster.
\end{abstract}

\maketitle
\section{Introduction}
A variety of cluster states have been known in light nuclei, 
such as $\alpha$+$\alpha$ and $^{16}$O+$\alpha$ 
states in $^8$Be and $^{20}$Ne and $3\alpha$ states in $^{12}$C. Cluster motion in these cluster states 
has been theoretically investigated in details 
with such microscopic cluster models as 
the resonating group method (RGM) \cite{RGM} and the generator coordinate method (GCM) \cite{GCM1,GCM2}
(for example, see Ref.~\cite{Fujiwara-supp} and references therein).
In this decade, a new interpretation of the cluster states has been proposed in Refs.~\cite{Tohsaki:2001an,Funaki:2002fn,Funaki:2003af,Funaki:2009zz,Zhou:2012zz,Zhou:2013ala}.
That is the dilute cluster gas state where clusters are not localized but they are rather 
freely moving occupying the lowest orbit in the cluster mean-field potential. 

One of the typical examples is the $3\alpha$ cluster state of $^{12}$C($0^+_2$). A new method of cluster model 
has been constructed for treating the $\alpha$ cluster gas state 
originally based on the spherical 
Gaussian. 
Nowadays, it is called "Tohsaki-Horiuchi-Schuck-R\"opke wave function" (THSR) \cite{Tohsaki:2001an}. 
The spherical THSR wave function has been extended to the deformed version \cite{Funaki:2002fn,Funaki:2003af}, 
and it has been shown that, when the $J^\pi$-projection and the orthogonality to the $^{12}$C($0^+_1$)
are taken into account, the single deformed THSR 
wave function is in principle equivalent to 
the full solution of the $3\alpha$ wave function obtained by
RGM and GCM calculations. In this paper, we call the 
spherical and deformed versions of 
THSR, "sTHSR" and "dTHSR", respectively. 
Also in the case of $^8$Be$(0^+_1$), the exact solution of the $2\alpha$ state obtained by the GCM calculation
can be described almost perfectly by the single
$2\alpha$ dTHSR wave function.

For the study of $^{20}$Ne,  Zhou {\it et al.} have introduced the generalized THSR wave function 
to investigate the cluster structure in $K^\pi=0^-_1$ band as well as $K^\pi=0^+_1$ band
and have shown that the single THSR wave functions give the better description than the single 
Bink-Bloch (BB) wave function \cite{brink66} having the localized Gaussian form with the fixed range parameter.
In particular, for $^{20}$Ne($1^-$), the single dTHSR is almost equivalent to the exact solution of the 
GCM with 99.98\% squared overlap.  
The fact that the single TSHR wave functions give better description than the single BB wave function indicates that 
the interpretation of the localized $^{16}$O+$\alpha$ cluster for 
the inversion doublet of $K^\pi=0^+_1$ and $K^\pi=0^-_1$ bands in $^{20}$Ne 
is too simple, but the quantum fluctuation of $\alpha$ cluster position is significant
as already known in the success of the GCM calculation with the 
superposition of many BB wave functions \cite{Nemoto72}.
In the works  with the THSR wave functions,   
the delocalization of clusters in $^{20}$Ne has been stressed and explained as the new 
concept of the "nonlocalization" \cite{Zhou:2012zz,Zhou:2013ala}.

Based on the success in the description of $2\alpha$, $3\alpha$, and $^{16}$O+$\alpha$ systems
with a single THSR wave function, the container picture has been recently proposed 
to understand the cluster states \cite{Zhou:2013eca}. 
In the container picture, clusters are moving 
in the lowest orbit of 
the cluster mean-filed potential whose spatial size is specified by the Gaussian range parameter 
of the THSR wave function.

However, one should be careful to discuss physical meaning of cluster wave function,  
in particular, in the inner region
where the cluster wave function is strongly affected by the antisymmetrization effect between 
clusters.
Moreover, the physical or mathematical meaning of the deformation of the Gaussian wave function
in the dTHSR is not obvious, because the angular momentum $l$-wave 
relative wave function projected from the deformed Gaussian shows a behavior quite different 
from a Gaussian function when the deformation is large. 

As already known, in the cluster states such as $\alpha$+$\alpha$ 
and $^{16}$O+$\alpha$ states in $^8$Be and $^{20}$Ne, 
the relative wave function between clusters 
is characterized by the suppressed inner part, 
the enhanced surface peak, and the outer tail.
The inner suppression and the enhanced amplitude at the surface can be understood by the strong 
antisymmetization effect between clusters. It is important that 
the inner nodal structure and the surface peak structure are dominantly determined by the 
antisymmetrization, and therefore, it is difficult to discuss the physical meaning of the 
original $\alpha$-cluster wave function before the antisymmetrization. 
On the other hand, the outer tail part is almost free from the antisymmetrization effect,
and it directly shows the $\alpha$ cluster motion in the physical state. 

The outer tail is caused by the quantum penetration and
its asymptotic behavior is well defined. Needless to say, 
the quantum penetration is important in particular 
in the loosely bound $\alpha$-cluster
system with the small $\alpha$ separation energy and the small 
centrifugal and Coulomb barriers.
In such a case, the wave function is slowly damping in the outer region and it has the remarkably long
outer tail. As a result,
the outer long tail becomes more significant, and hence, 
the description of the slowly damping tail part is essential for 
good description of the $\alpha$-cluster wave function. 

In that sense, it is clear that the BB wave function fails to describe the outer long tail
because it has the localized Gaussian form with the fixed range. 
The Gaussian function may give the better description if the Gaussian range is the adjustable parameter
as in the case of the sTHSR wave function. However, 
as mentioned above, the damping behavior of the tail part in the asymptotic region 
is well defined by the $\alpha$ separation energy as well as the 
centrifugal and Coulomb barriers, and obviously, it should be different from the 
Gaussian tail. One of the questions is why the 
projected dTHSR wave function can describe the correct damping behavior of the tail part 
and succeed to reproduce the exact solution almost perfectly.

In this paper, we investigate the $\alpha$-cluster motion 
in the $2\alpha$ and $^{16}$O+$\alpha$ 
cluster states of $^8$Be and $^{20}$Ne. 
The spinless two-body cluster systems of 
$^{16}$O+$\alpha$ and $\alpha$+$\alpha$
can be reduced to the one dimension problem with the relative coordinate $r$ between clusters. 
To discuss the physical feature of $\alpha$ cluster motion
we analyze the antisymmetrized relative wave function 
in one coordinate $r$ space, 
and discuss its behavior in three 
regions of $r$, the inner part, the surface peak, and the outer tail. 
In the present work, we consider several kinds of trial functions 
specified by one or at most two adjustable parameters and examine 
how accurately the trial function can reproduce the relative wave function of the exact solution
obtained by the GCM calculation. 
Particular attention should be paid on the detailed behavior of the outer tail 
to discuss the delocalization of the $\alpha$ cluster 
in weakly bound cluster states.

We consider the BB, sTHSR, dTHSR, and YT functions as trial functions. 
The relative wave functions in the BB, sTHSR, dTHSR, and YT functions are
characterized by the localized Gaussian with the fixed range, the spherical Gaussian, 
the deformed Gaussian, and the Yukawa tail function, respectively. 
Comparing the squared overlap of those trial functions with the GCM wave function,  
we discuss the quality of the description of the exact solution 
with those trial functions.
We show that the dTHSR is a good trial function which can 
give almost 100\% overlap with the exact solution of
weakly bound $\alpha$-cluster states such as $^8$Be($0^+$) and $^{20}$Ne($1^-$)
because the projected deformed Gaussian can fit the Yukawa-like tail in the outer region fairly well 
if the effective barrier hight in the outer region is low enough. 
We also demonstrate that 
a kind of wave function with a Yukawa tail gives a good description of those states
with almost the same quality as the dTHSR, and it gives even better description for such states 
as $^{20}$Ne($0^+$) and $^{20}$Ne($2^+_1$).

This paper is organized as follows. 
In the next section, the GCM calculation of $^8$Be  and $^{20}$Ne is explained. 
In \ref{sec:trial-functions}, the adopted trial functions
are described. 
The analyses of cluster states in 
$^8$Be and $^{20}$Ne are given in \ref{sec:8Be}
and \ref{sec:20Ne}, respectively.  Finally in \ref{sec:summary},
the discussion and summary are given. 
The tail behavior of the relative wave function in the dTHSR wave function 
in the large deformation limit is explained in the appendix.

%%%%%%%%%%%%%%%%%%%%%%%%%%%%%%%%%%%%%%%%%%%%%%%%%
\section{GCM calculation of $^8$Be and $^{20}$Ne}\label{sec:GCM}
%%%%%%%%%%%%%%%%%%%%%%%%%%%%%%%%%%%%%%%%%%%%%%%%%
The $\alpha+\alpha$ and $^{16}{\rm O}+\alpha$ cluster models 
are applied to $^8$Be and  $^{20}$Ne, respectively. To describe 
details of the $\alpha$ cluster motion, we solve the two-cluster problem 
with the GCM using BB cluster wave functions. 
In the GCM framework, we can obtain the precise relative wave function 
between the $\alpha$ cluster and the other cluster
by superposing BB wave functions.

\subsection{Brink-Bloch $\alpha$-cluster wave function and GCM}

We briefly review the BB wave function and the GCM calculation for 
a system composed of two spinless clusters $C_1$ and $C_2$. 
The mass numbers of $C_1$ and $C_2$ are $A_1$ and $A_2$ and the proton numbers are
 $Z_1$ and $Z_2$, respectively.
In the present case, $C_i$ is the $\alpha$ cluster or $^{16}$O.
In the GCM calculation of the $C_1$+$C_2$ cluster model, 
the total wave function 
can be expressed by the linear combination of 
BB wave functions \cite{brink66}.

A BB wave function of the two-cluster $C_1$+$C_2$ system
 with the relative position $\bvec{S}$ is expressed as 
\begin{eqnarray}
&&|\Phi_{\rm BB}(\bvec{S})\rangle = |\frac{1}{\sqrt{A!}}{\cal A} \{ \psi(C_1,\frac{-A_2}{A}\bvec{S}) \psi(C_2,\frac{A_1}{A}\bvec{S})\} \rangle.
\end{eqnarray}
Here $\psi(C_i,\bvec{S}_i)$ is the wave function of the $C_i$ cluster localized around $\bvec{S}_i$,
and it is given by the harmonic oscillator (H.O.) shell model wave function with the shifted center at 
$\bvec{S}_i$. ${\cal A}$ is the antisymmetrizer for all nucleons. 
$A=A_1+A_2$ is the total mass number.
The same H.O. width is chosen for $C_1$ and $C_2$ for simplicity.
We set the relative position $\bvec{S}$ on the $z$-axis $\bvec{S}=(0,0,S)$ in the intrinsic frame
and project the BB wave function to the spin-parity $J^\pi$ eigen state
\begin{eqnarray}
|\Phi^{J\pi}_{\rm BB}(\bvec{S}) \rangle &\equiv& N_l(S)
P^{J\pi}_{00} |\Phi_{\rm BB}(\bvec{S})\rangle,
%P^{J\pi}_{MK}&\equiv&P^\pi P^J_{MK}\\
%P^{\pi=\pm}&=&\frac{1\pm P_r}{2}\\
%P^J_{MK}&=&\frac{2J+1}{8\pi^2}\int d\Omega D^{J*}_{MK}(\Omega)R(\Omega).
\end{eqnarray}
where $P^{J\pi}_{MK}$ is the spin-parity projection operator, 
and $K=M=0$ is considered here because the BB wave function 
of two spinless clusters with $\bvec{S}=(0,0,S)$ is the $K=0$ eigen state.
The normalization factor $N_l(S)$ is chosen to be 
$N_l(S)=1/\sqrt{\langle\Phi_{\rm BB}(\bvec{S})|P^{J\pi}_{00} P^{J\pi}_{00}|\Phi_{\rm BB}(\bvec{S})\rangle}$ 
to satisfy the normalization $\langle \Phi^{J\pi}_{\rm BB}(\bvec{S})|\Phi^{J\pi}_{\rm BB}(\bvec{S}) \rangle=1$.

The GCM wave function for the $J^\pi$ state 
is given by the linear combination of the projected BB wave functions, 
\begin{eqnarray}\label{eq:gcm-BB}
|\Phi_{\rm GCM}\rangle &=&\sum_k c_k |\Phi^{J\pi}_{\rm BB}(\bvec{S}_k)\rangle.
\end{eqnarray}
Coefficients $c_k$ are determined by 
solving the discretized Hill-Wheeler equation which is equivalent to the diagonalization of
the norm and Hamiltonian matrices.
Here, the cluster-GCM wave function $\Phi_{\rm GCM}$ is normalized as
$\langle \Phi_{\rm GCM}|\Phi_{\rm GCM} \rangle=1$.

\subsection{Inter-cluster wave function and antisymmetrization effect}

In $|\Phi_{\rm BB}(\bvec{S})\rangle$, the relative wave function between clusters is written by a localized
Gaussian wave packet as,
\begin{eqnarray}
&&|\Phi_{\rm BB}(\bvec{S})\rangle = |\frac{1}{\sqrt{A!}}{\cal A} \{ 
\Gamma(\bvec{r},\bvec{S},\gamma) \phi(C_1)\phi(C_2) \phi_{\rm c.m.} \}
\rangle, \\
&&\Gamma(\bvec{S},\gamma;\bvec{r})= \left ( \frac{2\gamma}{\pi}\right )^{3/4}
e^{-\gamma(\bvec{r}-\bvec{S})^2} ,\\
&& \gamma \equiv  \frac{A_1A_2}{A}\frac{1}{2b^2},\\
&&\phi_{\rm c.m.}=\left( \frac{A}{\pi b^2} \right) e^{-\frac{A}{2b^2}\bvec{r}_G^2}.
\end{eqnarray}
$\bvec{r}$ is the relative coordinate between mass centers of clusters, 
$\phi(C_1)$ and $\phi(C_2)$ are internal wave functions of clusters,
$\bvec{r}_G$ is the coordinate of the center of the total mass, and 
$\phi_{\rm c.m.}$ is the wave function of the center of total mass motion.
$b$ is the width parameter of the H.O. for two clusters.

With the partial wave expansion of $\Gamma(\bvec{S},\gamma;\bvec{r})$, 
the $J^\pi$-projected BB wave function $|\Phi^{J\pi}_{\rm BB}(\bvec{S})\rangle$ for $\bvec{S}=(0,0,S)$
is rewritten,
\begin{eqnarray} \label{eq:chi}
&& |\Phi^{J\pi}_{\rm BB}(\bvec{S})\rangle= 
 |\frac{1}{\sqrt{A!}}{\cal A} \{ \chi^{\rm BB}_l(S;r)
Y_{l0}(\hat{r}) \phi(C_1)\phi(C_2) \phi_{\rm cm} \}\rangle, \\
&& \chi^{\rm BB}_l(S;r)=N_l(S)\sqrt{\frac{2l+1}{4\pi}}\Gamma_l(S,\gamma;r),\\
&& \Gamma_l(S,\gamma;r) \equiv 4\pi (\frac{2\gamma}{\pi})^{\frac{3}{4}} i_l(2\gamma S r) e^{-\gamma(r^2+S^2)},\\
\end{eqnarray}
where $i_l$ is the modified spherical Bessel function. $l$ equals to $J$ because two clusters are spinless. 
$\chi^{\rm BB}_l(S;r)$ is the radial part of the 
$l$-wave relative wave function in $|\Phi^{J\pi}_{\rm BB}(\bvec{S})\rangle$ before 
the antisymmetrization.

In the GCM wave function, the radial part $\chi^{\rm GCM}_l(r)$ of the $l$-wave 
relative wave function is given by the linear combination of $\chi^{\rm BB}_l(S;r)$, 
\begin{eqnarray}
&&|\Phi_{\rm GCM}\rangle=\sum_k c_k |\Phi^{J\pi}_{\rm BB}(\bvec{S}_k)\rangle =| \frac{1}{\sqrt{A!}} \mathcal{A} \left[ \chi^{\rm GCM}_l(r) Y_{l0}(\hat{r}) \phi(C_1)\phi(C_2)\phi_{\rm c.m.}  \right]  \rangle \\
&&\chi^{\rm GCM}_l(r)=\sum_k c_k  \chi^{\rm BB}_l(S_k;r) = \sum_k c_k  \sqrt{\frac{2l+1}{4\pi}} 
\Gamma_l(S_k,\gamma,r).
\end{eqnarray}
It means that the relative wave function $\chi_l(r)$ in the general $C_1$+$C_2$ cluster wave function
\begin{eqnarray}
|\Phi \rangle= |\frac{1}{\sqrt{A!}} \mathcal{A} \left[ \chi_l(r) Y_{l0}(\hat{r}) \phi(C_1)\phi(C_2) \phi_{\rm c.m.} 
\right]  \rangle,
\end{eqnarray}
is represented by the expansion of 
the function $\Gamma_l(S_k,\gamma;r)$ with various $S_k$ values in the GCM framework, 
and the coefficients are determined so as to minimize the energy of $|\Phi \rangle$.

The cluster wave function $\chi_l(r)$ before the antisymmetrization usually 
contains Pauli forbidden states of the inter-cluster motion
which vanish after the antisymmetrization of nucleons between $C_1$ and $C_2$ clusters.
Such forbidden states have no physical meaning in the total $C_1$+$C_2$ system, and therefore, 
in discussion of $\alpha$ cluster wave functions in physical states
we should extract physical component of the cluster wave function by eliminating 
unphysical forbidden states. 
For this aim, we use the antizymmetrized relative wave function $u_l(r)$ defined as follows, 
\begin{eqnarray}
\chi_l(r)&=&\sum_n a_n R_{nl}(b_r;r),\\
a_n&=& \int r^2dr R_{nl}(b_r;r) \chi_l(r),\\
u_l(r)&=&\sum_n  a_n \sqrt{\mu_{nl}} R_{nl}(b_r;r), \label{eq:ul}
\end{eqnarray}
where $R_{nl}(b_r;r)$ is the radial wave functions 
of H.O. with the width parameter $b_r=1/\sqrt{2\gamma}=\sqrt{A/A_1A_2}b$ and $\mu_{nl}$ is the eigen value of the RGM norm kernel \cite{Ikeda77-supp}.
$u_l(r)$ does not contain forbidden states, and it is normalized 
for the normalized total wave function $\langle\Phi|\Phi \rangle=1$ 
as,
\begin{eqnarray}
\int |u_l(r)|^2 r^2 dr=1.
\end{eqnarray}
It should be noted that $u_l(r)$ is not the so-called reduced width amplitude 
as it is defined by the weight $\sqrt{\mu_{nl}}$ instead of the weight $\mu_{nl}$. 
We regard the function $u_l(r)$ as the physical relative wave function, i.e., 
the radial part of inter-cluster wave function in the physical component, 
because it is free from forbidden states and it satisfies the normalization which is essential for the interpretation of the 
$\alpha$ cluster probability.
Moreover, the squared overlap between 
two normalized wave functions $|\Phi\rangle$ and $|\Phi'\rangle$
for a $C_1$+$C_2$ cluster system equals to the squared 
overlap between 
the antisymmetrized relative wave functions $u_l(r)$ and $u'_l(r)$ for $|\Phi\rangle$ and $|\Phi'\rangle$,
\begin{eqnarray}
|\langle \Phi|\Phi'\rangle|^2 = |\langle u_l(r) | u'_l(r) \rangle|^2. \label{eq:norm-over}
\end{eqnarray}
Here we define 
\begin{eqnarray}
\langle f(r) | g(r) \rangle\equiv \int f^*(r) g(r) r^2 dr
\end{eqnarray}
for given functions $f(r)$ and $g(r)$. Then, if we have the exact wave function $|\Phi\rangle$ and an approximated wave function 
$|\Phi'\rangle$, the accuracy of the approximated wave function 
can be judged by the squared overlap between the relative 
wave functions 
$u_l(r)$ and $u'_l(r)$ for $|\Phi\rangle$ and $|\Phi'\rangle$.

The original function $\chi_l(r)$ before the antisymmetrization 
and the corresponding antisymmetrized relative wave function 
$u_l(r)$ have 
the same asymptotic behavior in the large $r$ region where 
the antisymmetrization effect between clusters vanishes while they are different 
in the inner region where clusters largely overlap with each other 
and the relative wave function $u_l(r)$ is strongly 
affected by the antisymmetrization effect.

%\subsection{Inter-cluster wave functions }

\section{Descriptions of inter-cluster wave function} \label{sec:trial-functions}

\subsection{Description with trial functions and tail behavior}

By performing the GCM calculation with enough number basis wave functions given by BB wave functions, 
we obtained the $C_1$+$C_2$ cluster wave functions 
for $J^\pi$ states of $^8$Be and $^{20}$Ne. 
The obtained GCM wave function 
$|\Phi_{\rm GCM}\rangle$ is considered to be the exact solution in the full model space of
$C_1$+$C_2$ clusters.

In the GCM calculation, the wave function is expressed by the linear combination of BB wave functions. 
The superposition of THSR wave functions proposed in Refs.~\cite{Tohsaki:2001an,Funaki:2003af,Zhou:2012zz,Zhou:2013ala} 
is an alternative choice of basis wave functions. 
For a spinless two-body cluster system, which can be reduced to one-dimensional problem, 
the superposition of THSR wave functions is 
equivalent to that of BB wave functions because both of them cover the full model space of 
$C_1$+$C_2$ cluster states in principle. 
Namely, the relative wave function is expressed 
by the linear combination of shifted Gaussian 
functions with the fixed range 
in the former case, and it is given by 
the linear combination of various range Gaussians (the multirange Gaussian)
around the origin in the latter case. 

If the cluster state can be approximated well by a single basis wave function, 
one may consider that the basis wave function reflects important character of the 
inter-cluster motion. 
In Refs.~\cite{Funaki:2002fn,Funaki:2009zz,Zhou:2012zz,Zhou:2013ala}, it was shown that the single dTHSR wave function gives 
pretty good description of $\alpha$ cluster states of $^8$Be and $^{20}$Ne 
rather than the single BB wave function, and the container picture was proposed that the $\alpha$ cluster is delocalized 
distributing whole the system.

However, one should take care about the physical meaning of 
cluster wave function because the original relative wave 
function $\chi_l(r)$ before the antisymmetrization 
contains unphysical forbidden states. Usually, 
the microscopic wave function of the total system 
is not sensitive to the inner part of $\chi_l(r)$ because of the strong antisymmetrization effect. 
Therefore, it is difficult to judge which trial function
is best for the inner region.
Physical properties of $\alpha$ cluster states are characterized by 
the enhanced $\alpha$-cluster probability at the surface and the $\alpha$-cluster tail 
in the outer region of the physical relative wave function $u_l(r)$ after the antisymmetrization. The outer tail originates in the quantum penetration through the effective barrier and 
its asymptotic behavior is well defined by the $\alpha$ separation energy $|E_r|$ ($E_r$ is the energy measured from 
the $\alpha$ threshold energy).
In the ideal cluster system without cluster breaking, 
low-$l$ states with the small separation energy 
and the low Coulomb barrier should have the enhanced $\alpha$-cluster tail, which can be interpreted 
as almost zero-energy free $\alpha$ gas.
It means that the delocalization of the $\alpha$ cluster in the outer region
is an obvious consequence of the quantum penetration in the weakly bound low-$l$ $\alpha$ cluster
state in light systems such as $^{20}$Ne and $^8$Be 
with the low Coulomb barrier.

To discuss the feature of $\alpha$ cluster motion
in the physical region, we analyze the 
antisymmetrized relative wave function $u_l(r)$ and discuss its behavior in three 
parts, the inner part, the surface peak, and the outer tail.
We examine the features of $u_l(r)$ and also those of the 
original function $\chi_l(r)$ before the antisymmetrization of 
trial functions.
The inner part of $u_l(r)$ is less sensitive to the original relative wave function 
$\chi_l(r)$ because of the strong antisymmetrization effect,
while the outer tail of $u_l(r)$ directly reflects the tail behavior of the original function $\chi_l(r)$. 
Particular attention should be paid on the tail behavior 
to discuss the delocalization of $\alpha$ cluster in weakly bound cluster states.

As already mentioned, spinless two-body cluster systems such as 
$^{16}$O+$\alpha$ and $\alpha$+$\alpha$
can be reduced to the one dimension problem to describe the relative 
wave function $\chi_l(r)$. 
The squared overlap of the total wave function is given by the squared overlap of the antisymmetrized 
relative wave function $u_l(r)$ as Eq.~\ref{eq:norm-over}.
In the present work, we consider several kinds of trial functions for $\chi_l(a_1,a_2;r)$ 
specified by one or at most two adjustable parameters $a_1$ and $a_2$ and examine 
how accurately the trial function can reproduce the exact solution $u^{\rm GCM}_l(r)$.
For the criterion of the accuracy, we adopt the squared overlap,
\begin{equation}
{\cal N}^{\rm over}(a_1,a_2)\equiv |\langle  u_l(a_1,a_2;r)|u^{\rm GCM}_l(r) \rangle |^2.
\end{equation} 
If the total wave function given by the trial model function 
$\chi_l(a_1,a_2;r)$ is 100\% equivalent to the exact GCM wave function, the corresponding $u_l(a_1,a_2;r)$ 
has 100\% overlap with the exact solution $u^{\rm GCM}_l(r)$ as 
${\cal N}^{\rm over}(a_1,a_2)=1$.
Based on the criterion, we determine the optimum 
parameters $a_1$ and $a_2$ for the trial model wave function 
solving the mathematical problem to maximize the squared overlap
${\cal N}^{\rm over}(a_1,a_2)$. 
With the maximum value ${\cal N}^{\rm over}_{\rm max}$
of the squared overlap, we can discuss 
the accuracy of the model wave functions. 

Since the delocalization of the $\alpha$ cluster in the weakly bound cluster state
is characterized by the outer tail, the description of the tail behavior of $u^{\rm GCM}_l(r)$ with the trial model function
is crucial in the accurate reproduction of 
the exact solution.
For the quantitative discussion of the tail behavior, 
we analyze the curvature ${\cal C}_r(r)$ 
of $ru_l(r)$ which corresponds to the 
radial term of the local kinetic energy defined as, 
\begin{equation}
{\cal C}_r(r)\equiv\frac{\hbar^2}{2\mu} \frac{1}{r u_l(r)} \frac{d^2 (r u_l(r))}{dr^2}.
\label{eq:C}
\end{equation}
We call ${\cal C}_r(r)$ the radial curvature. 
In the enough large $r$ region free from the 
nucleus-nucleus potential, ${\cal C}_r(r)$ for the exact 
wave function $ru^{\rm GCM}_l(r)$ approaches the asymptotic solution, 
\begin{equation}\label{eq:C-asymp}
{\cal C}_r(r)=
\frac{\hbar^2}{2\mu}\frac{l(l+1)}{r^2} + Z_1 Z_2 \frac{e^2}{r} - E_{\rm r}
\end{equation}
When we omit the $r$-dependence of the Coulomb potential and consider the $l=0$ bound state, ${\cal C}_r(r)$ is approximately 
constant in the asymptotic region
and the tail part of $u^{\rm GCM}_l(r)$ is given
by the Yukawa function $e^{-\kappa r}/r$. 
More generally, 
the value of the radial curvature ${\cal C}_r(r)$ in the outer region relates to the local damping factor of the tail, 
that is,  small (large) ${\cal C}_r(r)$ means the slow (rapid) 
damping of the $\alpha$-cluster tail.

The delocalization of the $\alpha$ cluster originates in the 
quantum penetration though the effectively low barrier and it is characterized by 
the slowly damping long tail. In good reproduction of  
the exact solution for the weakly bound cluster state 
with a trial function, the
original trial function $\chi_l(a_1,a_2;r)$ before the antisymmetrization 
should be able to describe 
the correct tail behavior of the exact solution. 
By analyzing the radial curvature ${\cal C}_r(r)$ 
of $r\chi_l(r)$ and $ru_l(r)$ before and after the antisymmetrization, respectively, we show how well trial functions 
can describe the tail part of the exact $\alpha$-cluster wave function.

For the trial functions, we consider a function projected from 
a shifted spherical Gaussian function and that from 
a deformed Gaussian around the origin. The former is a 
model function 
which contains relative wave functions of
the BB wave function and the sTHSR wave function. 
The latter, the deformed Gaussian, corresponds to the 
dTHSR wave function.
We also consider a trial function with a Yukawa tail (YT).
In the later sections, we analyze relative wave functions as well as the radial curvature
of trial functions comparing those of the exact solution. 
We explain the reason why 
the dTHSR wave function 
gives good description than the BB wave function
for $^8$Be and $^{20}$Ne.
We also demonstrate that the YT function can well reproduce the exact solution equivalently to 
or even better than the dTHSR wave function.

\subsection{Shifted spherical Gaussian}
As a trial function for the inter-cluster wave function $\chi(\bvec{r})$
in the $C_1$+$C_2$ cluster wave function
\begin{equation}
|\Psi\rangle = |\frac{1}{\sqrt{A!}}{\cal A} \{ 
\chi(\bvec{r}) \phi(C_1)\phi(C_2) \}\rangle, 
\end{equation}
we adopt the shifted spherical Gaussian (ssG)
\begin{equation}
\chi(\bvec{r})=e^{-\frac{(\bvec{r}-\bvec{S})^2}{\sigma^2}}.
\end{equation}
The Guassian center position $\bvec{S}$ is chosen to be $\bvec{S}=(0,0,S)$. 
Then using the partial wave expansion of $\chi(\bvec{r})$,
the $l$-wave relative wave function of the $J^\pi=l^{(-1)^l}$ state projected from $\Psi$ 
is written as 
\begin{equation}
\chi^{\rm ssG}_l(S,\sigma;r)\propto 
i_l(\frac{2Sr}{\sigma^2}) e^{-\frac{r^2+S^2}{\sigma^2}}.
\end{equation}
For $\chi^{\rm ssG}_l(S,\sigma;r)$, the antisymmetrized relative 
wave function $u^{\rm ssG}_l(S,\sigma;r)$ is defined by the relation \ref{eq:ul}.
The normalization is chosen to be $\langle u^{\rm ssG}_l(r)|u^{\rm ssG}_l(r)\rangle =1$.

The shifted spherical Gaussian is parametrized by $S$ for the Gaussian center position 
and $\sigma$ for the Gaussian range. 
The parameters $S$ and $\sigma$ are optimized so as to maximize
the squared overlap $|\langle u^{\rm ssG}_l(S,\sigma;r)|u^{\rm GCM}_l(r)\rangle|^2$. 
The wave function is equivalent to the spherical case 
$\beta_x=\beta_y=\beta_z\equiv \beta$ of the hybrid
THSR wave function proposed in Ref.~\cite{Zhou:2013ala}, and present parameters
correspond to $S=S_z$ and $\sigma=\sqrt{A/A_1A_2}\sqrt{b^2+2\beta^2}$ with the parameters
$S_z$ and $\beta$ defined in Ref.~\cite{Zhou:2013ala}. 

\subsubsection{Brink-Bloch wave function}
The BB wave function corresponds to the shifted spherical 
Gaussian with the fixed Gaussian range $\sigma_{\rm fix}=1/\sqrt{\gamma}=
\sqrt{A/A_1A_2}b$. 
$\chi^{\rm BB}_l(S;r)$ and $u^{\rm BB}_l(S;r)$ equal to $\chi^{\rm ssG}_l(S,\sigma_{\rm fix};r)$
and $u^{\rm ssG}_l(S,\sigma_{\rm fix};r)$, respectively. 
The relative wave function in the BB wave function is specified 
by the parameter $S$ for the Gaussian center position. The parameter $S$ for the 
optimum BB wave function is determined so as to 
maximize the squared overlap $|\langle u^{\rm BB}_l(S;r)|u^{\rm GCM}_l(r)\rangle|^2$. 

$\sigma_{\rm fix}$ is comparable to or even smaller 
than the $\alpha$ cluster size $b$. Since the width is fixed to be $\sigma_{\rm fix}$, the relative wave function 
$\chi^{\rm BB}_l(r)$ of the BB wave function is localized around $r=S$. 
The radial curvature 
${\cal C}_r(r)$ of $r\chi^{\rm BB}_l(r)$ is roughly estimated to be 
\begin{equation}
{\cal C}_r(r)\approx \frac{\hbar^2}{\mu}\left(\frac{2(r-S)^2}{\sigma^4_{\rm fix}}
-\frac{3}{\sigma^2_{\rm fix}}+\frac{2S}{r}\right),
\end{equation}
approximating $\chi^{\rm BB}_l(r)$ by the Gaussian function
$\exp(-\frac{({r}-{S})^2}{\sigma_{\rm fix}^2})$ 
because it is the function $l$-projected from 
$\exp(-\frac{(\bvec{r}-\bvec{S})^2}{\sigma^2})$.
In the tail region $r>s$, ${\cal C}_r(r)$ increases rapidly reflecting the rapid damping tail of $\chi^{\rm BB}_l(r)$ because of the small range $\sigma_{\rm fix}$. 
It is clear that the BB function is not suitable to describe a slowly damping tail. 

\subsubsection{sTHSR wave function: $r^l$-weighted spherical Gaussian function}
When we take the $S\to 0$ limit of the shifted spherical Gaussian, 
the relative function goes to the $r^l$-weighted Gaussian around the origin, 
\begin{equation}
\lim_{s\to 0} \chi^{\rm ssG}_l(S,\sigma;r)\propto r^l e^{-\frac{r^2}{\sigma^2}}.
\end{equation}

For even $l$ states, this is equivalent to the spherical limit $\beta_\perp\rightarrow \beta_z$ $(\beta_x=\beta_y=\beta_\perp)$
of the deformed THSR wave function used for Be and Ne in Refs.~\cite{Funaki:2002fn,Zhou:2012zz}. 
For odd $l$ states, it is the spherical case of the zero limit intercluster distance parameter 
of the hybrid THSR wave function in Ref.~\cite{Zhou:2013ala}. 
We call this trial function the "spherical THSR" (sTHSR) in this paper, 
\begin{equation}
\chi^{\rm sTHSR}_l(\sigma;r)\propto r^l e^{-\frac{r^2}{\sigma^2}}.
\end{equation}
The sTHSR wave function is parametrized by the Gaussian range $\sigma$. 
Strictly speaking, $\sigma$ should be $\sigma\ge\sqrt{2A/A_1A_2}b$
in the sTHSR wave function because of the correspondence 
$\sigma^2=2A(b^2+2\beta^2)/A_1A_2$. 
In the present work, 
$\sigma$ is optimized so as to maximize the squared overlap $|\langle u^{\rm sTHSR}_l(\sigma;r)|u^{\rm GCM}_l(r)\rangle|^2$ for the normalized wave function
as $\langle u^{\rm sTHSR}_l(\sigma;r)| u^{\rm sTHSR}_l(\sigma;r)\rangle=1$. 

The radial curvature ${\cal C}_r(r)$ of $r\chi^{\rm sTHSR}_l(\sigma;r)$ is
trivial because the $r^l$-weighted Gaussian is the lowest solution 
for the $l$ state in the H.O. potential, 
\begin{eqnarray}
&&{\cal C}_r(r)= \frac{\hbar^2}{\mu}\left(\frac{2r^2}{\sigma^4}
-\frac{2l+3}{\sigma^2}\right)+ \frac{\hbar^2}{2\mu} \frac{l(l+1)}{r^2} =  \frac{\hbar^2}{2\mu} \frac{l(l+1)}{r^2}-
\hbar\omega \left(l+\frac{3}{2}\right)+\frac{1}{2}\mu\omega^2 r^2, \label{eq:curvature-sTHSR} \\
&&\omega=\frac{2\hbar}{\mu\sigma^2}.
\end{eqnarray}
In the outer region where the $1/r^2$ term is negligible and the $r^2$ term is dominant, 
${\cal C}_r(r)$ increases quadratically and crosses
the ${\cal C}_r(r)=0$ line around $r=\sqrt{l+\frac{3}{2}}\sigma$.
Because the width $\sigma$ is the adjustable parameter, 
the sTHSR can be a better function to describe the outer tail of the relative wave function 
than the BB wave function with the fixed range $\sigma_{\rm fix}$. However, since its curvature ${\cal C}_r(r)$ 
contains 
the quadratic term, it is difficult to perfectly reproduce a Yukawa-like long tail. 
Namely, the damping behavior of Gaussian tail is inconsistent with the Yukawa-like tail.
It is the mathematically consequence of Gaussian function. 
It indicates that some improvement is necessary in the tail part of trial functions
for better agreement to the exact solution.

\subsection{Deformed Gaussian function: deformed THSR wave function}

Another extension of the Gaussian function is the axial symmetric deformed Gaussian (dG) function
around the origin,  
\begin{eqnarray}
\chi^{\rm dG}(\sigma_\perp,\sigma_z;\bvec{r})&\propto &\exp(-\frac{x^2}{\sigma^2_\perp}-\frac{y^2}{\sigma^2_\perp}
-\frac{z^2}{\sigma^2_z})\\
&=&\exp \left( -\frac{r^2}{\sigma^2_\perp}+\frac{r^2}{\Delta}\cos \theta^2 \right) \\
\frac{1}{\Delta} &\equiv& \frac{1}{\sigma^2_\perp}-\frac{1}{\sigma^2_z}.
\end{eqnarray}
The relative wave function $\chi_l(r)$ of the even $l$ wave is given as 
\begin{eqnarray}
\chi^{\rm dG}_l(\sigma_\perp,\sigma_z;r)&\propto & \int Y_{l0}(\hat{\bvec{r}})\chi^{dG}(\bvec{r}) d\Omega \nonumber \\
& = & \sqrt{(2l+1)\pi} \int^\pi_0 \exp \left( -\frac{r^2}{\sigma^2_\perp}+\frac{r^2}{\Delta}\cos \theta^2 \right)
P_l(\cos \theta) \sin \theta d\theta \nonumber \\
&=& 2 \sqrt{(2l+1)\pi} \exp \left( -\frac{r^2}{\sigma^2_\perp}\right)  
\int^1_{0} P_l(t) \exp \left(\frac{r^2}{\Delta}t^2 \right) dt, \label{eq:dGl}
\end{eqnarray}
where $P_l(t)$ is the Legendre polynomial. 

For odd $l$ state, we adopt the axial symmetric deformed Gaussian function infinitesimally 
shifted to $z$ direction from the origin, 
\begin{eqnarray}
\chi^{\rm dG-odd}(\sigma_\perp,\sigma_z;\bvec{r})&\propto &z\exp(-\frac{x^2}{\sigma^2_\perp}-\frac{y^2}{\sigma^2_\perp}
-\frac{z^2}{\sigma^2_z})\\
&=&r\cos \theta\exp \left( -\frac{r^2}{\sigma^2_\perp}+\frac{r^2}{\Delta}\cos \theta^2 \right).
\end{eqnarray}
The relative wave function $\chi_l(r)$ for the odd $l$ wave is given as 
\begin{eqnarray}
\chi^{\rm dG}_l(\sigma_\perp,\sigma_z;r)&\propto & \int Y_{l0}(\hat{\bvec{r}})\chi^{\rm dG-odd}(\bvec{r}) d\Omega \\
&=& 2 \sqrt{(2l+1)\pi} r \exp \left( -\frac{r^2}{\sigma^2_\perp}\right)  
\int^1_{0} P_l(t) t \exp \left(\frac{r^2}{\Delta}t^2 \right) dt,
\end{eqnarray}
where $P_l(t)$ is the Legendre polynomial. 
$\chi^{\rm dG}_l(\sigma_\perp,\sigma_z;r)$ is parametrized by the range parameters
$\sigma_\perp$ and $\sigma_z$ which are optimized so as to maximize 
the squared overlap $|\langle u^{\rm dG}_l(\sigma_\perp,\sigma_z;r)|u^{\rm GCM}_l(r)\rangle|^2$
for the normalized wave function
as $\langle u^{\rm dTHSR}_l(\sigma_\perp,\sigma_z;r)| u^{\rm dTHSR}_l(\sigma_\perp,\sigma_z;r)\rangle=1$. 

The present deformed Gaussian wave function for even $l$ states
corresponds to the deformed THSR proposed in Ref.~\cite{Funaki:2002fn}, 
and that for odd $l$ states corresponds to the zero limit of the 
intercluster distance parameter of the hybrid THSR proposed 
in Ref.~\cite{Zhou:2013ala}.
The parameters $\beta_{\perp,z}$ of the deformed THSR and the hybrid THSR relate to 
the present parameters $\sigma_{\perp, z}$ as 
$\sigma_\perp^2=2A(b^2+2\beta^2_\perp)/A_1A_2$ and $\sigma_z^2=2A(b^2+2\beta^2_z)/A_1A_2$. 
Because of these relations,  the parameters $\sigma_\perp$ and $\sigma_z$ should be
$\sigma_\perp\ge \sqrt{2A/A_1A_2}b$ and $\sigma_z\ge \sqrt{2A/A_1A_2}b$ in the deformed THSR.
When we impose this condition, $\sigma_\perp, \sigma \ge \sqrt{2A/A_1A_2}b$,
we call the deformed Gaussian wave function "deformed THSR" (dTHSR).
% and use the notation
%$\chi^{\rm dTHSR}_l(r)$ and $u^{\rm dTHSR}_l(r)$ for the relative wave functions
%before and after antisymmetrization.

In the spherical limit $\sigma_\perp \to \sigma_z$, 
$\chi^{\rm dG}_l(r)$ goes to the $S\to 0$ limit $\chi^{\rm ssG}_l(r)$,
and its radial curvature ${\cal C}_r(r)$ becomes the quadratic form given 
in Eq.~\ref{eq:curvature-sTHSR}.
On the other hand, in the case of the largely deformed Gaussian 
with $\sigma_z \gg \sigma_\perp$, the curvature $|{\cal C}_r(r)|$ of $r\chi^{\rm dG}_l(r)$ can be small 
in the outer region. As explained in the appendix, for $\chi^{\rm dG}_l(r)$ with 
$\sigma_z \gg \sigma_\perp$, 
${\cal C}_r(r)$ goes to 0 
in the $\sigma_\perp \ll r < \sigma_z$ region.
It means that the deformed Gaussian should be a better trial function that can 
efficiently describe the slow damping behavior of long tail than the spherical Gaussian having 
the Gaussian tail. 

\subsection{Yukawa tail function}

To describe the slow damping behavior of the $\alpha$-cluster tail in the outer region, 
we consider another trial function with a Yukawa tail in the outer region
instead of Gaussian functions.
To avoid the singularity of the Yukawa function in the small $r$ region, we
introduce a Yukawa tail (YT) function 
by smearing the inner part of the Yukawa function and continuously connecting 
to the $r^l$ function in the $r\to 0$ limit which is the correct asymptotic behavior at $r\to 0$ 
of the regular wave function in the finite potential well as follows,
\begin{eqnarray}
\chi^{\rm YT}_l(a_R,a_Y;r)&\propto & \left(\frac{r}{R(r)}\right)^l 
F^{\rm yukawa}\left(\frac{R(r)}{a_Y}\right)\\
F^{\rm yukawa}(x)&=&\frac{{\mathrm e}^{-x}}{x}\\
R(r)&=&\frac{r}{\sqrt{1-\exp(-r^2/a_R^2)}}.
\end{eqnarray}
Here $R(r)$ is the scaling function that approaches $R(r)=a_R$ in $r\to 0$ and
it goes to $R(r)=r$ in the large $r$ region. Therefore, the function 
$\chi^{\rm YT}_l(a_Y,a_R;r)$ has the Yukawa tail in the outer region.
The damping behavior of the tail is characterized by the parameter $a_Y$. 
The parameter $a_R$ corresponds to the range for smearing 
the Yukawa function in the inner region. 
To describe the Yukawa tail in the outer region,  
the smearing range $a_R$ should be the same order of or smaller than the 
surface peak position of the $\alpha$-cluster wave function.
For the optimum YT function, these two parameters $a_Y$ and $a_R$ are optimized 
so as to maximize the squared overlap $|\langle u^{\rm YT}_l(a_Y,a_R;r)|u^{\rm GCM}_l(r)\rangle|^2$
for the normalized wave function as 
$\langle u^{\rm YT}_l(a_Y,a_R;r) |\langle u^{\rm YT}_l(a_Y,a_R;r)\rangle=1$. 

%We also consider the YT function with the fixed $a_R$ parameter to be $a_R=\sqrt{2}b_r$
%which corresponds to the same range of the Gaussian of $R_{00}(b_r;r)$.
%Thus $a_R$-fixed YT function is specified by one parameter $a_Y$ for the Yukawa tail 
%which is optimized so as to the squared overlap.

It is clear that the curvature ${\cal C}_r(r)$ of
$r\chi^{\rm YT}_l(\sigma,a;r)$ goes to constant in the outer tail part, 
\begin{eqnarray}
{\cal C}_r(r)=\frac{\hbar^2}{2\mu}\frac{1}{a_Y^2}.
\end{eqnarray}

Instead of Yukawa function, 
we can also consider an alternative function having the tail of 
modified spherical Hankel function $H_l(r)$ as,
\begin{eqnarray}
\chi^{\rm HT}_l(a_R,a_Y;r)&\propto & \left(\frac{r}{R(r)}\right)^lH_l\left(\frac{R(r)}{a_Y}\right),
\end{eqnarray}
which has the curvature in the outer tail,
\begin{equation}
{\cal C}_r(r)=\frac{\hbar^2}{2\mu}\frac{l(l+1)}{r^2}+\frac{\hbar^2}{2\mu}\frac{1}{a_Y^2}.
\end{equation} 
We checked this trial function 
and found that it gives almost the same quality as the YT function in 
fitting the exact solution $u^{\rm GCM}_l(r)$. 
In this paper, we show only the results of the YT function. 

\section{Results of $^{8}$Be} \label{sec:8Be}

\subsection{GCM calculation of $^{8}$Be}
For $^8$Be, we perform the GCM calculation of the $2\alpha$ cluster model. 
We use the Volkov No.1 with $m=0.56$ and the width parameter $b=1.36$ fm
the same as the $2\alpha$ calculation in Ref.~\cite{Funaki:2002fn}. 
$S_k=0.5,1.25,2.0,\cdots,20.0$ fm are chosen for the basis BB wave functions in the GCM calculation.
In the present calculation, two-body Coulomb force is approximated by the seven-range Gaussian.
The calculated energy $E_{\rm r}$ of $^{8}$Be($0^+_1$) measured from the $2\alpha$ 
threshold energy is  $E_{\rm r}=-0.32$ MeV which slightly underestimates 
the experimental energy $E_{\rm r}=0.092 MeV$. 

\subsection{Squared overlap of trial functions with the GCM wave function of $^8$Be}

We consider how well trial functions can describe the exact solution obtained by 
the GCM calculation. 
Trial functions are specified by one of two parameters. 
We vary the parameter(s) and search for the optimum parameter(s) 
which gives the maximum value of the squared overlap 
${\cal N}^{\rm over}=|\langle u_l(r)|u^{\rm GCM}_l(r)\rangle|^2$ with the GCM wave function. 
For trial functions, we adopt the BB, sTHSR, dTHSR, and YT wave functions.
The maximum values ${\cal N}^{\rm over}_{\rm max}$ of the squared overlap ${\cal N}^{\rm over}$
and the optimized parameters 
are shown in table \ref{tab:8Be-over}.
The squared overlap of $R_{nl}(b_r;r)$ for the relative wave function 
of the SU(3) shell model (SM) limit is also shown. 

The results for the BB, sTHSR, dTHSR wave functions are in principle the same as those 
discussed by Funaki et al. in Refs.~\cite{Funaki:2002fn,Funaki:2009zz}. 
The description of the BB wave function is worse compared with the sTHSR wave function although it is much better than the SM wave function. 
The reason is that the BB wave function can describe the enhanced surface peak 
of the antisymmetrized relative wave function $u^{\rm GCM}_l(r)$ in $^8$Be$(0^+)$
better than the SM one but it fails to describe the long outer tail.
The sTHSR wave function can describe the outer tail part better than the BB wave function 
but it is not sufficient for 
the perfect description because a Gaussian tail is different
from the correct asymptotic behavior of the relative wave function.  
On the other hand, the dTHSR wave function describes the GCM wave function 
almost perfectly as 99.99$\%$ squared overlap. 
This is because the outer tail behavior is described fairly well by the dTHSR
as shown in Ref.~\cite{Funaki:2009zz}.

It is found that the present YT wave function gives almost equal quality to 
the dTHSR in describing the GCM wave function. It indicates that 
the Yukawa-like tail is essential to reproduce the exact solution.
%It is interesting that the $a_R$-fixed YT function has 99.97\% overlap and gives
%better result than the sTHSR wave function 
%even though it has only one adjustable parameter indicating that 
%a Yukawa tail is considered to be better description than a Gaussian tail. 

\begin{figure}[tb]
\begin{center}
	\includegraphics[width=7cm]{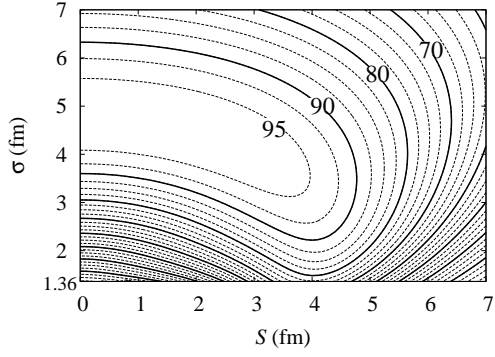} 	
\end{center}
\vspace{0.5cm}
  \caption{\label{fig:il-norm}
Squared overlap ${\cal N}^{\rm over}=|\langle u_l(r)|u^{\rm GCM}_l(r)\rangle|^2$ 
for the ssG wave function $u^{\rm ssG}_l(S,\sigma;r)$ as functions of 
the distance parameter $S$ and the range parameter $\sigma$. 
The data on the horizontal axis at $\sigma=1.36$ fm correspond to 
${\cal N}^{\rm over}$ for the BB wave function with a given $S$ value, 
while those on the vertical axis at $S=0$ 
are ${\cal N}^{\rm over}$ for the sTHSR function with a given $\sigma$ value. 
Solid lines indicate contour for 90\%, 80\%, 70\%, $\cdots$. Between the solid lines,
thin dotted lines are drawn at 2.5\% intervals.  
}
\end{figure}

The main reason for the failure of the BB wave function in describing the GCM wave function 
is that the long tail is missing because the range $\sigma$ of the shifted spherical Gaussian 
is fixed to be the small value $\sigma_{\rm fix}=1.36$ fm in the BB wave function.
In Fig.~\ref{fig:il-norm}, we show the squared overlap 
${\cal N}^{\rm over}=|\langle u_l(r)|u^{\rm GCM}_l(r)\rangle|^2$ for the ssG wave function 
$u^{\rm ssG}_l(S,\sigma;r)$ as functions of the parameters $S$ and $\sigma$. 
The values on the $\sigma=1.36$ fm line correspond to 
${\cal N}^{\rm over}$ for the BB wave function with a given $S$ value, 
while those on the $S=0$ line are ${\cal N}^{\rm over}$ for the sTHSR function 
with a given $\sigma$ value. 
There exists a wilde plateau with ${\cal N}^{\rm over}\ge 95\%$ 
in the region $\sigma=4\sim 5$ fm and $S=0\sim 4$ fm. 
Even if the parameter $S$ is fixed to $S=4$ fm, 
the ssG function can have about $95\%$ overlap with the GCM wave function 
by adjusting $\sigma$. The optimized $\sigma$ for $S=4$ fm is $\sigma=3.5$ fm 
which is much larger than the $\sigma_{\rm fix}$ in the BB wave function
and can describe the outer tail of the GCM wave function reasonably.
The maximum ${\cal N}^{\rm over}$ is given at $S=0$ which is contained in the 
model space of the sTHSR function.
This behavior is consistent with the argument for the delocalization of $\alpha$ cluster 
in Refs.~\cite{Funaki:2009zz,Zhou:2013ala}. 
It should be noted that the inner part of $\chi_l(r)$ 
is less significant because the physical wave function is not so sensitive 
to the inner part of the original function $\chi_l(r)$ before the antisymmetrization, 
while the tail part of $\chi_l(r)$ is important relatively. 
In the optimized ssG wave function,
the $S=0$ is chosen for the best fit of the slow damping feature of 
the outer long tail within the ssG model space.
Nevertheless, the wide plateau from $S=0$ fm to $S\sim 4$ fm at $\sigma=4\sim 5$ fm
indicates less importance of the inner part but the particular 
importance of the long outer tail which should be regarded as 
the delocalization of cluster because of the quantum penetration.

\begin{table}[ht]
\caption{Maximum values ${\cal N}^{\rm over}_{\rm max}$ ($\%$) of the squared overlap 
${\cal N}^{\rm over}=|\langle u_l(r)|u^{\rm GCM}_l(r)\rangle|^2$ and the optimized parameters (fm) for 
the trial functions, BB, sTHSR, dTHSR, and  YT wave functions for $^8$Be($0^+_1$).
The squared overlap ${\cal N}^{\rm over}$ of 
$R_{20}(b_r;r)$ for the SU(3) shell model limit 
with $u^{\rm GCM}_l(r)$ is also shown. 
\label{tab:8Be-over}}
\begin{center}
\begin{tabular}{cccccc}
\hline
\hline
    & BB & sTHSR & dTHSR & YT & SM \\
    & ${\cal N}^{\rm over}_{\rm max}(S)$ &
    ${\cal N}^{\rm over}_{\rm max}(\sigma)$ &
    ${\cal N}^{\rm over}_{\rm max}(\sigma_\perp,\sigma_z)$ &
   
    ${\cal N}^{\rm over}_{\rm max}(a_R,a_Y)$ & ${\cal N}^{\rm over}$ \\
\hline
$^{8}$Be($0^+_1$) & 77.27(4.01) & 97.29(4.77) & 99.99(2.88,11.06) 
 & 99.98(2.02,3.27)& 21.79 \\ 
\hline
\hline
\end{tabular}
\end{center}
\end{table}

\subsection{Analysis of $\alpha$-$\alpha$ intercluster wave functions}

We analyze the antisymmetrized relative wave functions $u_l(r)$ as well as 
the non-antisymmetrized ones $\chi_l(r)$ of the trial functions comparing them 
with the exact solution $u^{\rm GCM}_l(r)$ obtained by the GCM calculation. 

\begin{figure}[tb]
\begin{center}
	\includegraphics[width=13.5cm]{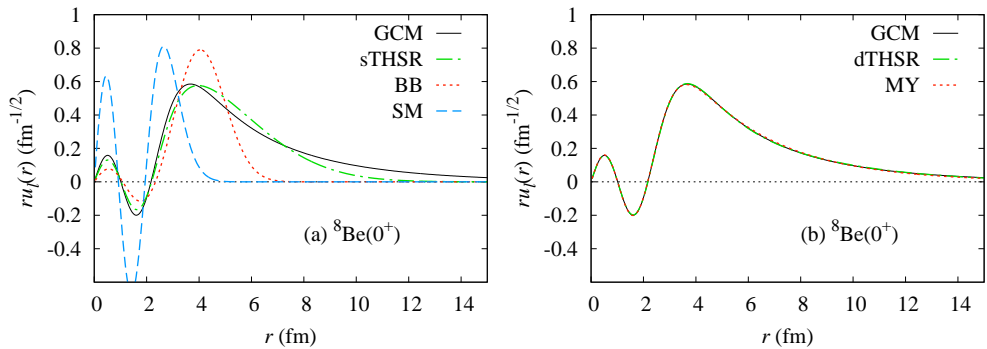} 	
\end{center}
\vspace{0.5cm}
  \caption{\label{fig:be8-ru}
Relative wave functions $ru_l(r)$ of the optimized trial wave functions for $^8$Be($0^+_1$) 
compared with that of the GCM wave function.  
$rR_{20}(b_r;r)$ for the SU(3) shell model limit is also shown. 
}
\end{figure}

The antisymmetrized relative wave functions $ru_l(r)$ of the optimized trial wave functions 
are shown in Fig.~\ref{fig:be8-ru}. The relative wave function 
for the SU(3) SM limit given by the function $rR_{20}(b_r;r)$ 
is also shown in the figure for comparison.
$ru^{\rm GCM}_l(r)$ of the GCM wave function is characterized by
three parts, the inner part, the surface peak, and the outer tail.
The inner part has the oscillating nodal structure because of the antisymmetrization effect. 
Compared with the shell model limit case, 
the inner part is suppressed, while the surface peak is enhanced and shifted toward outward 
in $ru^{\rm GCM}_l(r)$. Moreover, $ru^{\rm GCM}_l(r)$ has the long tail in the outer region
because of the quantum penetration. The tail part is damping very slowly 
because of the small separation energy $|E_r|=0.32$ MeV in $^8$Be($0^+_1$). As a result, 
the tail component contributes to the significant fraction of the total probability
$|\langle u^{\rm GCM}_l(r)|u^{\rm GCM}_l(r)\rangle|=1$.

\begin{figure}[tb]
\begin{center}
	\includegraphics[width=13.5cm]{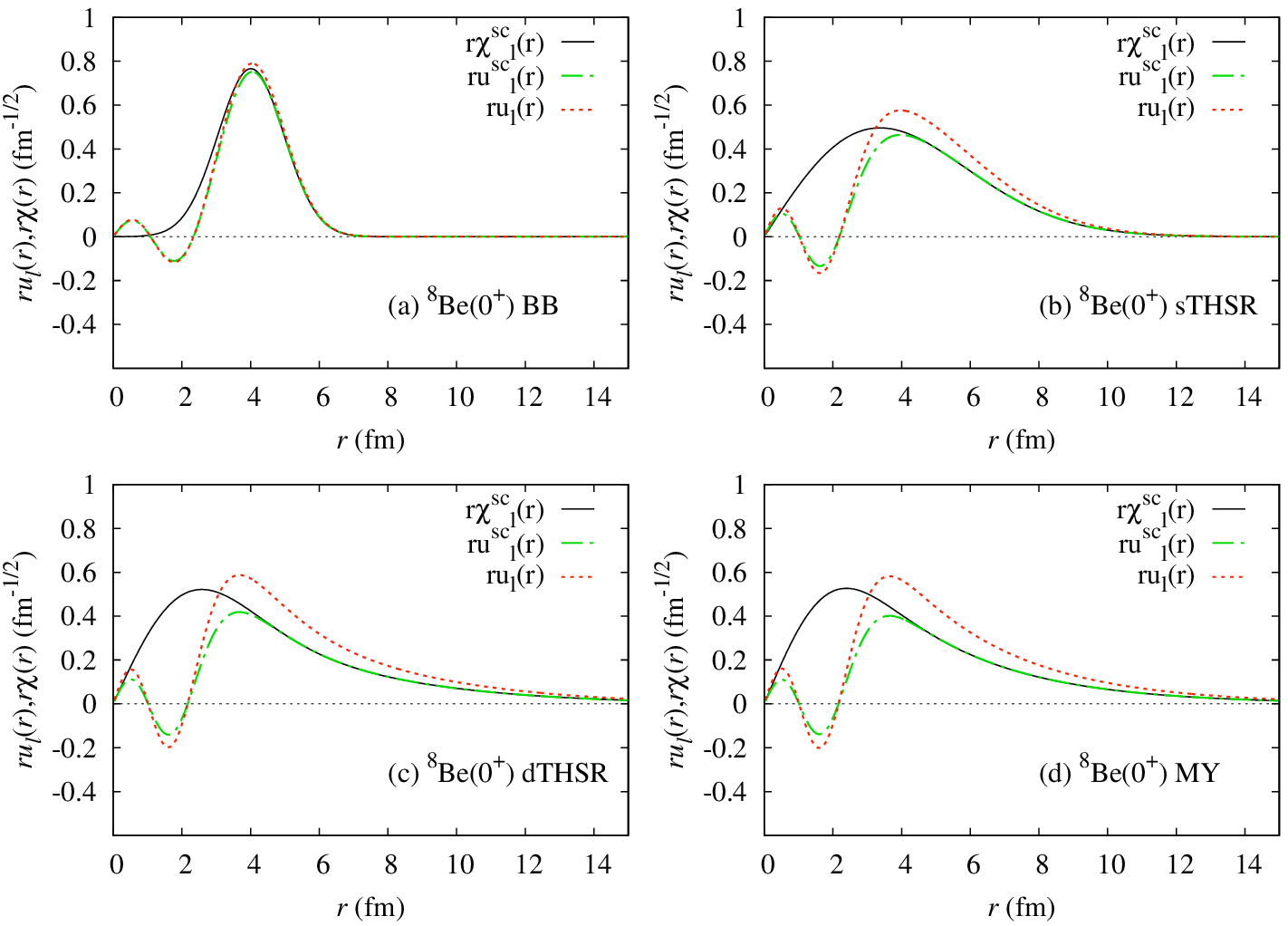} 	
\end{center}
\vspace{0.5cm}
  \caption{\label{fig:be8-chi-u}
Relative wave functions $r\chi_l(r)$ before 
the antisymmetrization and $ru_l(r)$ after the antisymmetrization
of the optimized trial functions for 
$^8$Be($0^+_1$). The scaled functions 
$r\chi^{\rm sc}_l(r)\equiv r\chi_l(r)/\sqrt{\langle \chi_l(r)|\chi_l(r)\rangle}$ and 
$ru^{\rm sc}_l(r)\equiv ru_l(r)\sqrt{\langle \chi_l(r)|\chi_l(r)\rangle}$ are shown 
as well as $ru_l(r)$.   
}
\end{figure}

The original relative wave functions $r\chi_l(r)$ before the antisymmetrization
of the optimized trial functions are shown compared with the antisymmetrized relative 
wave functions 
$ru_l(r)$ after the antisymmetrization in Fig.~\ref{fig:be8-chi-u}. 
In the figure, the scaled functions 
$r\chi^{\rm sc}_l(r)\equiv r\chi_l(r)/\sqrt{\langle \chi_l(r)|\chi_l(r)\rangle}$ and 
$ru^{\rm sc}_l(r)\equiv ru_l(r)\sqrt{\langle \chi_l(r)|\chi_l(r)\rangle}$ are shown 
as well as the normalized wave function $ru_l(r)$. 
The structures of the inner nodal oscillation and the surface peak in $ru_l(r)$ 
are not so sensitive to $r\chi_l(r)$ because of the antisymmetrization effect.
On the other hand, 
in the region $r>4$ fm, the $ru^{\rm sc}_l(r)$ almost 
agrees to $r\chi^{\rm sc}_l(r)$ because the antisymmetization effect vanishes
in the outer region. This means that the shape of the tail part of 
$ru_l(r)$ is directly determined by the shape the original function $r\chi_l(r)$.

In the comparison of the BB wave function with the GCM wave function, 
it is found that the surface peak structure as well as 
the suppressed inner part are described reasonably but the outer long tail is missing
in the BB wave function
because of the rapid damping of the Gaussian with the fixed range $\sigma_{\rm fix}$.
The description of the outer tail is drastically improved by the sTHSR wave function
in which the Gaussian range $\sigma$ is the adjustable parameter.
However, the Gaussian shape of the sTHSR wave function is insufficient to describe the detailed tail behavior.
The dTHSR and YT wave functions show almost the perfect agreement to $ru^{\rm GCM}_l(r)$. 
The success of the YT wave function indicates that the Yukawa-type tail
of the YT function is suitable to reproduce the outer long tail rather 
than the Gaussian tail.  

As shown in table ~\ref{tab:8Be-over}, the squared overlap ${\cal N}^{\rm over}$ 
of the trial functions with 
the GCM wave function is much small as about $20\%$ in the case of SU(3) shell model.
It remarkably increases in the BB wave function as ${\cal N}^{\rm over}_{\rm max}\sim 80\%$. 
It means that $80\%$ of the wave function can be reproduced 
by describing the inner suppression and the enhanced surface peak while 
the description of the outer tail behavior 
is essential for the remaining $20\%$ accuracy.

\begin{figure}[tb]
\begin{center}
	\includegraphics[width=7cm]{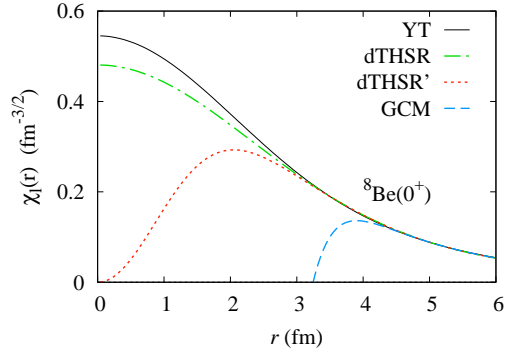} 	
\end{center}
\vspace{0.5cm}
  \caption{\label{fig:be8-chi-r}
Relative functions $\chi^{\rm sc}_l(r)$
of the GCM wave function, 
and the optimized dTHSR and YT wave functions for $^8$Be($0^+_1$).
The function $\chi_l(r)-(chi_l(0)/R_{00}(b_r;0)) R_{00}(b_r;r)$ (dTHSR') 
modified from $\chi_l(r)$ of the dTHSR function is also shown.
}
\end{figure}

The relative wave function $\chi_l(r)$ before 
the antisymmetrization contains unphysical forbidden states, and therefore, 
one should care about that the inner part of $\chi_l(r)$ has no or less physical meaning. 
Indeed, the physical relative wave function $u_l(r)$ after the antisymmetrization 
does not depend so much 
on the detailed behavior of the inner part of $\chi_l(r)$. In Fig.~\ref{fig:be8-chi-r},
we demonstrate that almost the same $u_l(r)$ can be obtained from 
various $\chi_l(r)$ having different inner structures.
In the figure, we show $\chi_l(r)$ for the dTHSR and YT wave functions. 
Although both of them give almost the same $u_l(r)$ functions almost equivalent to 
the exact solution $u^{\rm GCM}_l(r)$, some difference is shown 
in $\chi_l(r)$ in the $r<2$ fm region. 
We also show a modified relative wave function 
by subtracting $R_{00}(b_r;0)$ for the forbidden state from $\chi^{\rm dTHSR}_l(r)$ 
by hand as $\chi^{\rm dTHSR'}_l(r)\equiv \chi^{\rm dTHSR}_l(r)-(\chi^{\rm dTHSR}_l(0)/R_{00}(b_r;0)) R_{00}(b_r;r)$ (labeled dTHSR' in the figure).
The modified function gives the physical relative wave function 
$u_l(r)$ completely same as that 
of $\chi^{\rm dTHSR}_l(r)$. We also show the $\chi^{\rm GCM}_l(r)$ before the antisymmetrization
of the GCM wave function. 
It is found that the inner part of $\chi_l(r)$ 
does not affect the physical state. 

Thus, in the inner region 
it is in principle difficult to 
discuss the localization or delocalization of cluster 
because of the insensitivity to the original trial function. 
On the other hand, 
the cluster structure is characterized by the enhanced surface peak and the outer tail.
The localization and delocalization of cluster can be definitely distinguished
by the damping behavior of the outer tail, which is caused by the quantum penetration.
Because of the small $\alpha$ separation energy of 
$^8$Be($0^+_1$), $u^{\rm GCM}_l(r)$  has the slowly damping tail indicating the delocalization 
of cluster.

\begin{figure}[tb]
\begin{center}	
	\includegraphics[width=13.5cm]{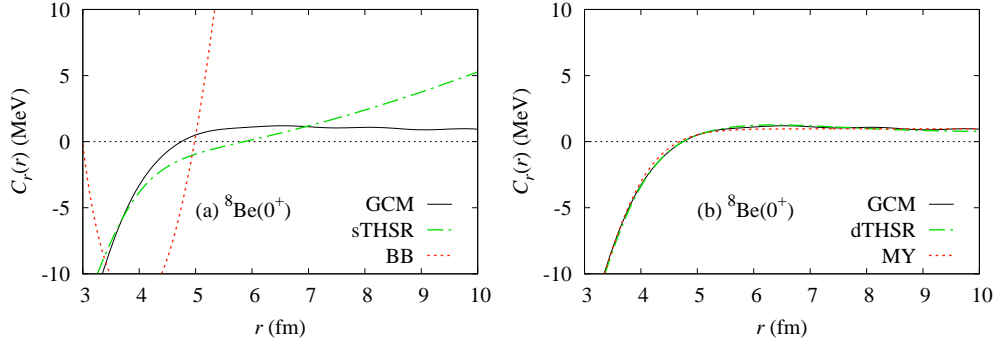} 	
\end{center}
\vspace{0.5cm}
  \caption{\label{fig:be8-VE}
Radial curvature ${\cal C}_r(r)$ of the antisymmertized relative wave function $ru_l(r)$ of the optimized trial 
functions for $^8$Be($0^+_1$) 
compared with that of the GCM wave function. 
}
\end{figure}

\begin{figure}[tb]
\begin{center}	
	\includegraphics[width=13.5cm]{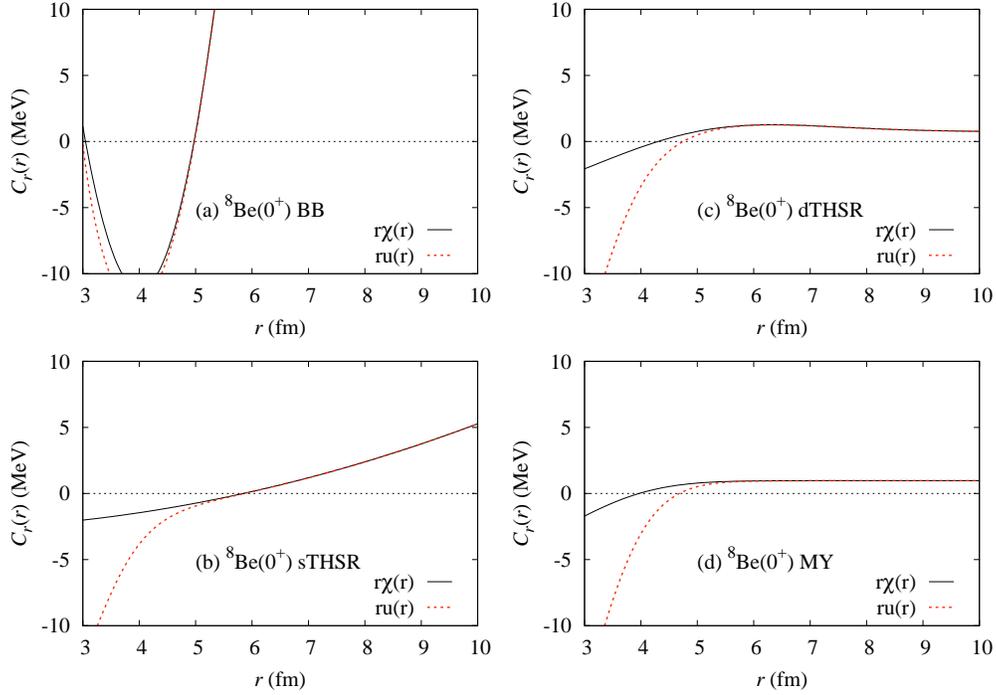} 	
\end{center}
\vspace{0.5cm}
  \caption{\label{fig:be8-VE-chi}
Radial curvature ${\cal C}_r(r)$ of the original relative wave function $r\chi_l(r)$ of the optimized trial 
functions for $^8$Be($0^+_1$)\.
}
\end{figure}

To discuss the damping behavior of the outer tail more quantitatively, 
we analyze the radial curvature ${\cal C}_r(r)$ of $ru_l(r)$ and $r\chi_l(r)$ defined 
in Eq.~\ref{eq:C}. 
In Fig.~\ref{fig:be8-VE}, the $r$ dependence of ${\cal C}_r(r)$ of $ru_l(r)$ for the
optimized trial function is compared with that for the GCM wave function. 
${\cal C}_r(r)$ of $r\chi_l(r)$ is also shown in Fig.~\ref{fig:be8-VE-chi}. 

As explained before,
in the outer region where the nucleus-nucleus interaction vanishes, 
the radial curvature 
${\cal C}_r(r)$ of the exact solution is well defined 
by the centrifugal barrier, the Coulomb barrier,
and the constant value $-E_r$ of the $\alpha$ separation energy
as given in Eq.~\ref{eq:C-asymp}. 
For the GCM wave function of $^{8}$Be($0^+_1$), ${\cal C}_r(r)$  is small
and almost flat in the the $r>5$ fm region
because of the small separation energy $|E_r|$ and the small Coulomb barrier and no centrifugal barrier. 
For trial functions, 
${\cal C}_r(r)$ of $ru_l(r)$ in $r>5$ fm region
directly reflects ${\cal C}_r(r)$ of the original $r\chi_l(r)$ 
which is simply given by the form of the model function. 
The BB wave function has the steep $r$ dependence of ${\cal C}_r(r)$ and can not describe 
the flat behavior of correct ${\cal C}_r(r)$ of the GCM wave function.
The sTHSR wave function gives better results
than the BB wave function. However, the ${\cal C}_r(r)$ of the Gaussian tail 
in the sTHSR is given by the Harmonic oscillator potential as Eq.~\ref{eq:curvature-sTHSR}, 
it is different from the flat behavior of the correct ${\cal C}_r(r)$.
On the other hand, the dTHSR and the YT wave functions 
can reproduce well the $r$ dependence of the correct ${\cal C}_r(r)$.
These results indicate that the slow damping of the outer tail of the 
GCM wave function is approximately 
described by the Yukawa tail rather than Gaussian tail.
The optimized dTHSR wave function mathematically has the slowly damping tail 
similar to the Yukawa tail.

\begin{figure}[tb]
\begin{center}	
	\includegraphics[width=13.5cm]{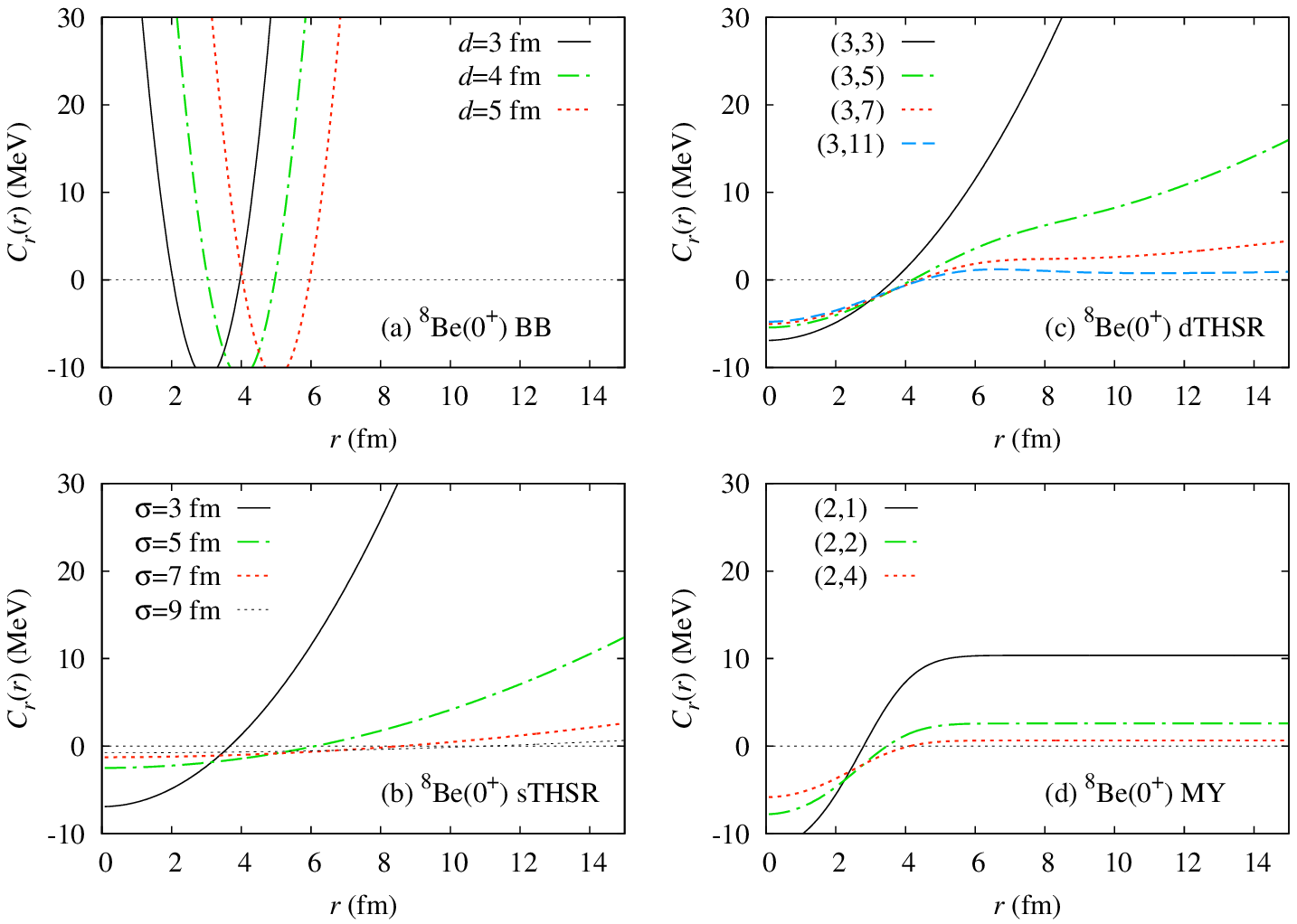} 	
\end{center}
\vspace{0.5cm}
  \caption{\label{fig:be8-VE-chi-parameter}
Radial curvature ${\cal C}_r(r)$ of $r\chi_l(r)$ of the BB, sTHSR, dTHSR, and YT functions 
with various parameters. 
(a) ${\cal C}_r(r)$ of the BB wave function with the parameters $S=3$, 4, 5 fm.
(b)  ${\cal C}_r(r)$ of the sTHSR wave function with the parameters $\sigma=3$, 5, 7, 9 fm.
(c)  ${\cal C}_r(r)$ of the dTHSR wave function with the parameters $(\sigma_\perp,\sigma_z)=(3,3)$, (3,5), (3,7), and (3,11) fm. 
(d)  ${\cal C}_r(r)$ of the YT wave function with the parameters $(a_R,a_Y)$=(2,1), (2,2), and (2,4) fm.  
}
\end{figure}

The $r$ dependence of the 
radial curvature of ${\cal C}_r(r)$ in the outer region is trivial 
for the BB, sTHSR, and YT wave functions as explained in the previous section,
while that for the dTHSR wave function is not trivial. 
We show, in Fig.~\ref{fig:be8-VE-chi-parameter}, ${\cal C}_r(r)$ of $r\chi_l(r)$ for 
trial functions with various parameter sets. 
In the BB wave function, 
the function ${\cal C}_r(r)$ is shifted with the change of the parameter $S$ while keeping 
the shape almost unchanged. 
In the case of the sTHSR wave function, ${\cal C}_r(r)$ is given by the H.O. potential 
with the frequency $\omega=2\hbar/\mu\sigma^2$ as shown in Eq.~\ref{eq:curvature-sTHSR}, 
The size of H.O., .i.e., the slope of ${\cal C}_r(r)$ depends on $\sigma$, and 
${\cal C}_r(r)$ crosses the 
${\cal C}_r(r)=0$ line around $r=\sqrt{l+3/2}\sigma$. 
Certainly, it is not be able to adjust the slope and the crossing point independently. 
As seen in Fig.~\ref{fig:be8-VE}, 
the optimized parameter $\sigma=4.77$ fm  shows the 
reasonable agreement to the correct ${\cal C}_r(r)$ in the $r=5\sim 8$ fm region.
However, ${\cal C}_r(r)$ of the sTHSR is the increasing function and fails to describe the
flat behavior even though the agreement is improved by the sTHSR with 
the large range parameter $\sigma$ than the BB wave function. 
${\cal C}_r(r)$ for the YT function is constant to be $\hbar^2/2\mu a^2_Y$
in the region roughly larger than $\sim 2a_R$. With the range parameter $a_Y$
of the Yukawa tail, the constant value of the flat region of ${\cal C}_r(r)$ 
can be freely adjusted. 
Indeed, with the optimized parameter $a_Y=3.27$ fm, ${\cal C}_r(r)$ of the GCM 
wave function in the tail region is reproduced well. 
 
The $r$ dependence of ${\cal C}_r(r)$ for the dTHSR is not trivial. 
In Fig.~\ref{fig:be8-VE-chi-parameter}, we show ${\cal C}_r(r)$ of the dTHSR of 
the prolately deformed case $\sigma_z >\sigma_\perp$ with the fixed 
$\sigma_\perp$. In the deformed case, 
the slope of ${\cal C}_r(r)$ changes around $r\sim 2\sigma_\perp$, and it is more gentle 
in the outer region than the inner region. 
As $\sigma_z$ increases, the gradient of ${\cal C}_r(r)$ in the 
outer region becomes small and ${\cal C}_r(r)$ approaches to the ${\cal C}_r(r)=0$ axis. 
For the optimized parameters $(\sigma_\perp,\sigma_z)=(2.88,11.06)$ fm having the large 
deformation, the ${\cal C}_r(r)$ in the outer region is quite different 
from that of the Gaussian tail.
Namely, ${\cal C}_r(r)$ is small and close to zero and it is in good agreement to the 
correct ${\cal C}_r(r)$ in the outer region. However, it should be noted that 
the  ${\cal C}_r(r)$ of the dTHSR wave function can describe the flat behavior only when 
the function is enough small as ${\cal C}_r(r)=0 \sim 1$ MeV, 
but it may fail to describe the flat function 
with a larger offset.
The $^8$Be($0^+_1$) system is the favorable case that 
the dTHSR wave function can fit the correct tail behavior of the 
GCM wave function having the small radial curvature ${\cal C}_r(r)$ 
because of the small $\alpha$ separation energy
as well as the small Coulomb barrier and no centrifugal barrier.
Such the slowly damping tail can be described well by the dTHSR wave function with the large 
deformation.

In the analysis of the relative wave function of $^8$Be, we can reach the following conclusions.
The relative wave function between two $\alpha$ clusters in $^8$Be system 
is characterized by three parts, the oscillating inner part, the enhanced surface peak, and 
the outer tail. 
The inner part is suppressed while the surface peak
is relatively enhanced. The nodal structure in the inner region and 
the enhanced peak structure at the surface in the physical wave function 
$ru_l(r)$ are not so sensitive to the original trial functions $\chi_l(r)$ 
because of the strong antisymmetrization effect between clusters. 
The outer tail is caused by the quantum penetration and its asymptotic behavior 
is well defined. Since $^8$Be($0^+_1$) is the weakly bound cluster state having 
the small $\alpha$ separation energy and the low Coulomb barrier and no centrifugal barrier, 
its wave function 
is slowly damping in the outer region, and the outer tail becomes the remarkably long tail.

To get a good approximation of the exact solution (GCM wave function) for 
such the weakly bound $\alpha$-cluster state,  
it is essential to fit the outer tail part, in particular, its slow damping behavior.
Since the inner and peak parts are mainly determined by the antisymmetrization effect and therefore it is relatively less important in the fitting.
The outer tail part is the slowly damping function characterized 
by the obvious effective barrier given by the Coulomb force and the separation energy 
free from the nuclear interaction, and therefore, 
it is interpreted as almost zero-energy free $\alpha$ gas.
This "free $\alpha$-cluster gas part" in the outer region can be understood as
the delocalization of cluster.
It should be pointed out that the origin of the delocalization, i.e.,  
the "free $\alpha$-cluster gas part" in the outer region, 
is the quantum penetration and it is the natural consequence of the weak binding of the 
$\alpha$ cluster. 
In the asymptotic region where the nucleus-nucleus interaction vanishes, 
the damping behavior of the outer tail in this free gas region is well defined by the 
$\alpha$ separation energy.
In the ideal case that the separation energy is
small and the Coulomb and centrifugal barriers are not high, 
the dTHSR wave function can describe the exact wave function
fairly well because it has the suitable form 
to fit the long tail in the outer free gas region. However, it should be stressed that
the success of the dTHSR is the mathematical result of the 
fact that the dTHSR gives the form different from the Gaussian tail but 
rather similar to the Yukawa-type tail. 

\section{Results of $^{20}$Ne}\label{sec:20Ne}

\subsection{GCM calculation of $^{20}$Ne}

For $^{20}$Ne, we perform the GCM calculation of the $^{16}$O+$\alpha$ cluster model. 
We use Volkov No.1 with $m=0.60$ and the width parameter $b=1.46$ fm
the same as the $^{16}$O+$\alpha$ calculations in Refs.~\cite{Zhou:2012zz,Zhou:2013ala}. 
$S_k=0.6,1.2,\cdots,12.0$ fm are chosen for the basis BB wave functions in the GCM calculation.
In the present calculation, two-body Coulomb force is approximated by the seven-range Gaussian.
For the $^{16}$O cluster wave function, we use the $4\alpha$ BB cluster wave function 
with an enough small intercluster distance which is equivalent to the $p$-shell closed H.O.
configuration. 

The energy of $0^+$, $2^+$, $4^+$, $6^+$, and $8^+$ states in the $K^\pi=0^+$ band
and that of $1^-$ and $3^-$ states in the $K^\pi=0^-$ band are shown in 
table \ref{tab:20Ne-ene}. 
The calculated $6^+$, $8^+$, and $3^-$ states are obtained within the 
bound state approximation in the present GCM basis $S_k\le 12$ fm
though they are resonances above the $\alpha$-decay threshold energy.
The present GCM calculation is in principle consistent with the "Brink GCM" calculation in 
Refs.~\cite{Zhou:2012zz,Zhou:2013ala}.

\begin{table}[ht]
\caption{Energy of $^{20}$Ne calculated with the cluster GCM using 
Volkov No.1 $m=0.60$ and $b=1.46$ fm. The energy $E_r$ (MeV) is measured from the $\alpha$ threshold energy. 
The experimental energy of the states in $K^\pi=0^+_1$ and $K^\pi=0^-_1$ bands is also listed. 
\label{tab:20Ne-ene}}
\begin{center}
\begin{tabular}{|c|rr|}
\hline
    & $E_{\rm r}$ & $E_{\rm r}$ \\
    & GCM & Exp. \\
    \hline
$^{20}$Ne($0^+_1$) & $-$5.92 & $-$4.73 \\
$^{20}$Ne($2^+_1$) & $-$4.71& $-$3.1 \\
$^{20}$Ne($4^+_1$) & $-$1.90& $-$0.48 \\
$^{20}$Ne($6^+_1$) & 2.55& 4.05 \\
$^{20}$Ne($8^+_1$) & 8.94& 7.22 \\
$^{20}$Ne($1^-_1$) & $-$1.24& 1.06 \\
$^{20}$Ne($3^-_1$) & 1.07 & 2.43 \\
\hline
\end{tabular}
\end{center}
\end{table}

\subsection{Squared overlap of trial functions with the GCM wave function of $^{20}$Ne}

In a similar way to the analysis of $^8$Be,  
we see how well trial functions can describe the exact solution of $^{\rm 16}$O+$\alpha$ 
cluster states obtained by the GCM calculation. 
For trial functions, we adopt the BB, sTHSR, dTHSR, and YT wave functions.
In addition we use the deformed Gaussian "dG" wave function which has 
the same form of the dTHSR wave function 
but no restriction of $\sigma_{\perp,z}\ge \sqrt{A/A_1A_2}b$ differently from the dTHSR wave function.

The maximum values ${\cal N}^{\rm over}_{\rm max}$ of the squared overlap 
${\cal N}^{\rm over}=|\langle u_l(r)|u^{\rm GCM}_l(r)\rangle|^2$ and the optimized parameters 
are shown in table \ref{tab:20Ne-over}.
The squared overlap of $u_l(r)=R_{nl}(b_r;r)$ in the SU(3) shell model limit 
with $u^{\rm GCM}_l(r)$ is also shown. Here the node number $n$ is 
$2n+l=8$ and $2n+l=9$ for even and odd $l$ states, respectively. 

The results for the BB and dTHSR wave functions are in principle the same as those discussed by 
Zhou {\it et al.} in Refs.~\cite{Zhou:2012zz,Zhou:2013ala}.
The  ${\cal N}^{\rm over}_{\rm max}$ of the BB wave function is much larger than ${\cal N}^{\rm over}$ of the SM 
wave function because it describes the enhanced surface peak better than the SM. 
However the description of the GCM wave function is as much as ${\cal N}^{\rm over}_{\rm max}=90\sim 95\%$ and it is not satisfactory 
because of the missing of the outer tail part in the BB wave function. The description is 
improved by the sTHSR wave function, in particular, for the weakly bound states such as $^{20}$Ne$(1^-_1)$ and 
$^{20}$Ne$(3^-_1)$.  Further improvement is given by the dTHSR wave function; ${\cal N}^{\rm over}_{\rm max}>99 \%$ is obtained 
by the dTHSR for $^{20}$Ne($0^+_1$), 
$^{20}$Ne($1^-_1$), and $^{20}$Ne($3^-_1$) as already shown in Ref.~\cite{Zhou:2013ala}. However, 
the description is not perfect for $J^\pi=2^+,4^+,6^+$, and $8^+$ states and maximum overlap is 
as much as ${\cal N}^{\rm over}_{\rm max}<99 \%$.

On the other hand, the YT function gives fairly good results for all states with more than 99\% accuracy except for $^{20}$Ne($6^+_1$).
This indicates that the YT function is the better trial function to fit the GCM wave function. 
It is interesting that the dG wave function without the restriction of $\sigma_{\perp,z}\ge \sqrt{A/A_1A_2}b$
shows the better result than the dTHSR except for $^{20}$Ne($1^-_1$). Namely, in the model space of the deformed Gaussian, 
the optimum solution exists in the $\sigma_{\perp}< \sqrt{A/A_1A_2}b$ or $\sigma_{z}< \sqrt{A/A_1A_2}b$ region 
out of the model space of the dTHSR. It is the mathematical results of the function projected from the deformed Gaussian
which favors the large deformation to fit the tail part of the GCM wave function.

\begin{table}[ht]
\caption{Maximum values ${\cal N}^{\rm over}_{\rm max}$ ($\%$) of the squared overlap 
${\cal N}^{\rm over}=|\langle u_l(r)|u^{\rm GCM}_l(r)\rangle|^2$ and optimized parameters (fm) for 
the trial functions of BB, sTHSR, dTHSR, dG, and  YT wave functions for $^{20}$Ne.
The squared overlap ${\cal N}^{\rm over}$ for
$R_{nl}(b_r;r)$ of the SU(3) shell model (SM) wave function 
with $u^{\rm GCM}_l(r)$ is also shown. 
\label{tab:20Ne-over}}
\begin{center}
\begin{tabular}{ccccccc}
\hline
\hline
    & BB & sTHSR & dTHSR & dG & YT & SM \\
    & ${\cal N}^{\rm over}_{\rm max}(S)$ &
    ${\cal N}^{\rm over}_{\rm max}(\sigma)$ &
    ${\cal N}^{\rm over}_{\rm max}(\sigma_\perp,\sigma_z)$ &
    ${\cal N}^{\rm over}_{\rm max}(\sigma_\perp,\sigma_z)$ &
    ${\cal N}^{\rm over}_{\rm max}(a_R,a_Y)$ & ${\cal N}^{\rm over}$ \\
\hline
$^{20}$Ne($0^+_1$)	& 93.50(3.24)	&	98.47(2.39)	&	99.29$(1.56	,3.00)$	&	99.67$(0.81	,3.02)$	&	99.94$(0.68	,0.83)$	&	41.04 \\
$^{20}$Ne($2^+_1$)	&	93.32(3.13)	&	97.46(2.04)	&	98.80$(1.15	,2.75)$	&	98.84$(1.02	,2.77)$	&	99.46$(1.40	,0.75)$	&	43.61 \\
$^{20}$Ne($4^+_1$)	&	93.16(2.87)	&	96.15(1.75)	&	97.84$(1.15	,2.39)$	&	99.25$(0.77	,2.71)$	&	99.79$(1.41	,0.76)$	&	50.08 \\
$^{20}$Ne($6^+_1$)	&	93.72(2.41)	&	95.33(1.50)	&  96.69$(1.15	,1.93)$ &	97.23$(1.00	,2.07)$	&	98.34$(2.04	,0.53)$	&	61.04 \\
$^{20}$Ne($8^+_1$)	&	95.65(1.73)	&	96.15(1.30)	&	98.45$(1.15	,1.87)$ &	98.45$(1.15	,1.87)$	&	99.57$(2.03	,0.55)$	&	76.21 \\
$^{20}$Ne($1^-_1$)	&	91.75(4.08)	&	99.53(2.97)	&	99.98$(4.31	,1.91)$	&	99.98$(4.31	,1.91)$	&	99.99$(2.60	,1.19)$	&	19.99 \\
$^{20}$Ne($3^-_1$)	&	89.87(4.03)	&	97.94(2.53)	&	99.85$(4.32	,1.15)$	&	99.90$(4.47	,0.87)$	&	99.98$(2.09	,1.39)$	&	20.43 \\
\hline
\hline
\end{tabular}
\end{center}
\end{table}

\subsection{Analysis of $^{16}$O-$\alpha$ intercluster wave functions}
\begin{figure}[tb]
\begin{center}	
	\includegraphics[width=13.5cm]{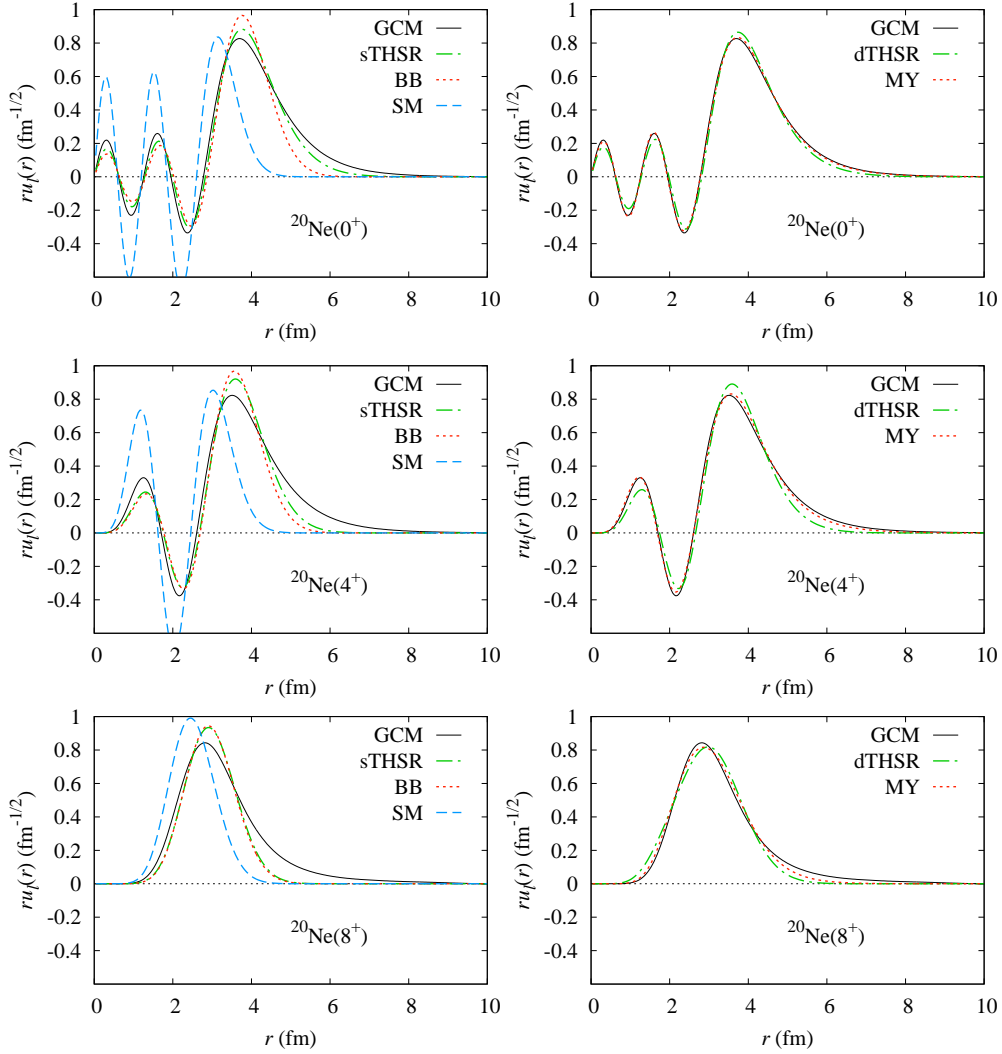} 	
\end{center}
\vspace{0.5cm}
  \caption{\label{fig:ne20-ru}
Relative wave functions $ru_l(r)$ of the optimized trial wave functions for $K=0^+_1$ band states of $^{20}$Ne
compared with that of the GCM wave function.  
$ru_l(r)=R_{nl}(b_r;r)$ for the SU(3) shell model limit is also shown. 
}
\end{figure}

\begin{figure}[tb]
\begin{center}	
	\includegraphics[width=13.5cm]{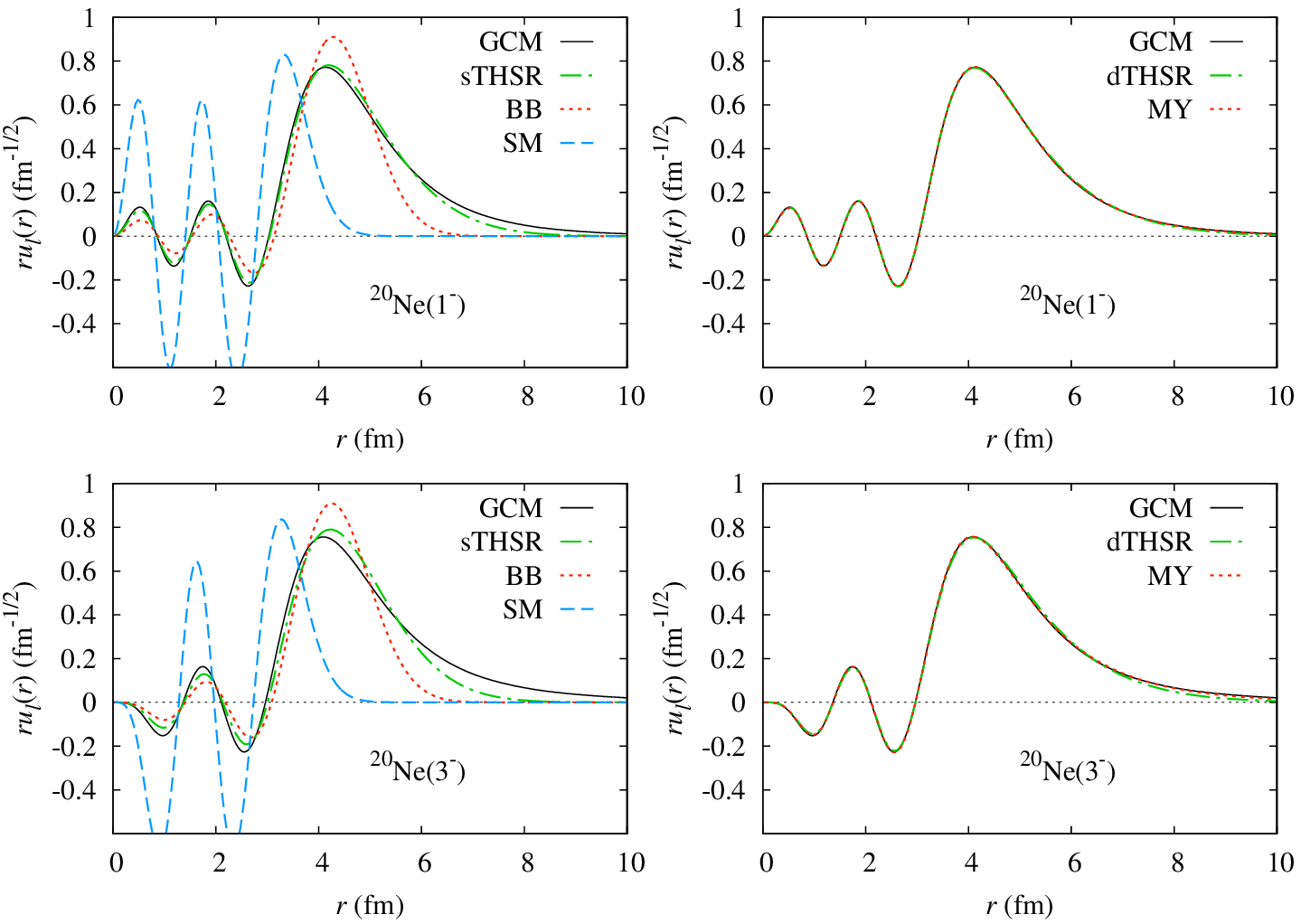} 	
\end{center}
\vspace{0.5cm}
  \caption{\label{fig:ne20-ru-np}
Same as Fig.~\ref{fig:ne20-ru} but for the $K=0^-_1$ band states of $^{20}$Ne.
}
\end{figure}

\begin{figure}[tb]
\begin{center}	
	\includegraphics[width=13.5cm]{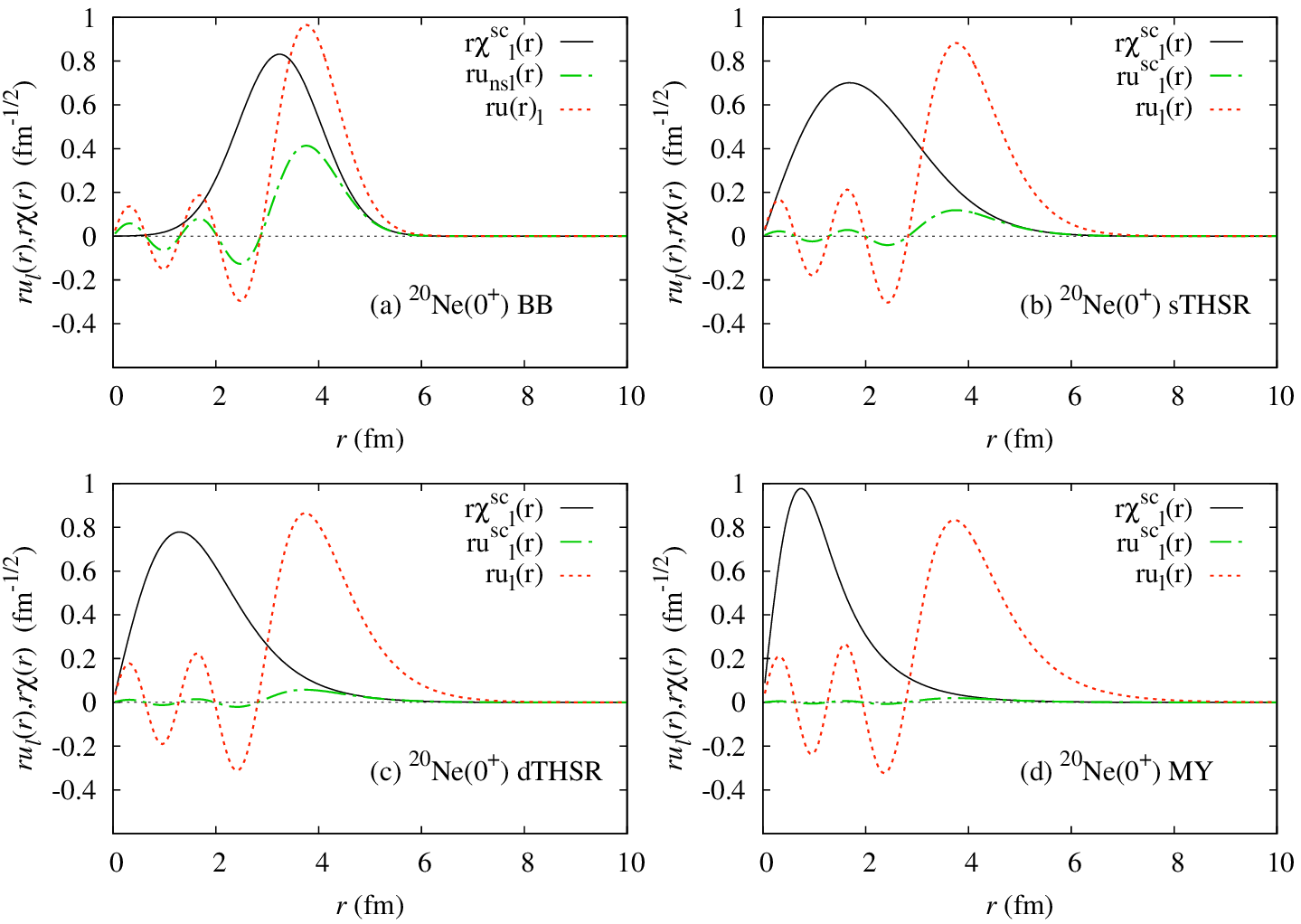} 	
\end{center}
\vspace{0.5cm}
  \caption{\label{fig:ne20-chi-u}
Relative wave functions $r\chi_l(r)$ before 
the antisymmetrization and $ru_l(r)$ after the antisymmetrization
of the optimized trial functions for 
$^{20}$Ne($0^+_1$). The scaled functions 
$r\chi^{\rm sc}_l(r)\equiv r\chi_l(r)/\sqrt{\langle \chi_l(r)|\chi_l(r)\rangle}$ and 
$ru^{\rm sc}_l(r)\equiv ru_l(r)\sqrt{\langle \chi_l(r)|\chi_l(r)\rangle}$ are shown 
as well as $ru_l(r)$.   
}
\end{figure}

In a similar way to the previous analysis of $^8$Be, we analyze the antisymmetrized relative wave function $u_l(r)$ 
as well as the non-antisymmetrized one $\chi_l(r)$ before the antisymmetrization. 
$ru_l(r)$ of the optimized trial functions is compared with the exact solution $ru^{\rm GCM}_l(r)$ 
in Figs.~\ref{fig:ne20-ru} and \ref{fig:ne20-ru-np}, and the relative wave function $r\chi_l(r)$ and $ru_l(r)$ before and after the antisymmetrization 
of the trial functions for $^{20}$Ne($0^+_1$) are shown in Fig.~\ref{fig:ne20-chi-u}. 
The inner oscillating part and the enhanced surface peak structures of $ru_l(r)$ 
are not so sensitive to the details of the original trial function $r\chi_l(r)$ at least in low $l$ states 
because of the strong antisymmetrization effect.  To describe well the exact solution with the trial function
it is essential to fit well the outer tail with $r\chi_l(r)$. 
Compared with $ru_l(r)$ of the SM wave function, the BB wave function 
describes well the suppressed inner nodal region and the enhanced surface peak but it fails to fit the outer tail
because of the fixed Gaussian range.
The sTHSR describes the outer tail somewhat better than the BB wave function, however, it is not so good in particular 
for positive-parity states in the $K^\pi=0^+_1$ band. The dTHSR wave function gives better results in the description of the outer tail than 
the sTHSR, especially fairly good description of the outer tail in $^{20}$Ne($1^-_1$).
However, it is still insufficient for positive-parity states. 
The best fit of the outer tail part of $ru^{\rm GCM}_l(r)$ is given by the YT function. 

\begin{figure}[tb]
\begin{center}	
	\includegraphics[width=13.5cm]{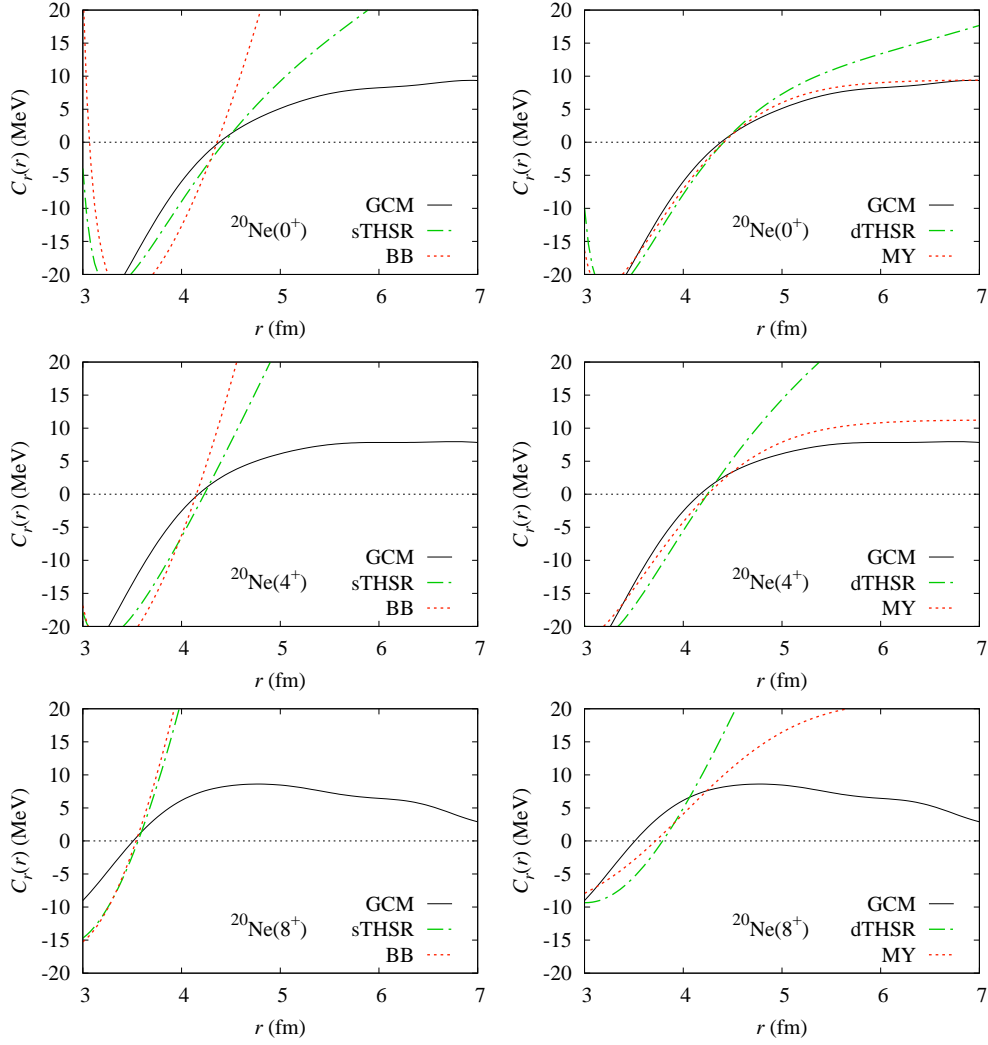} 	
\end{center}
\vspace{0.5cm}
  \caption{\label{fig:ne20-VE}
Radial curvature ${\cal C}_r(r)$ of $ru_l(r)$ of the optimized trial 
functions for the $K=0^+_1$ band states of $^{20}$Ne 
compared with that of the GCM wave function. 
}
\end{figure}

\begin{figure}[tb]
\begin{center}	
	\includegraphics[width=13.5cm]{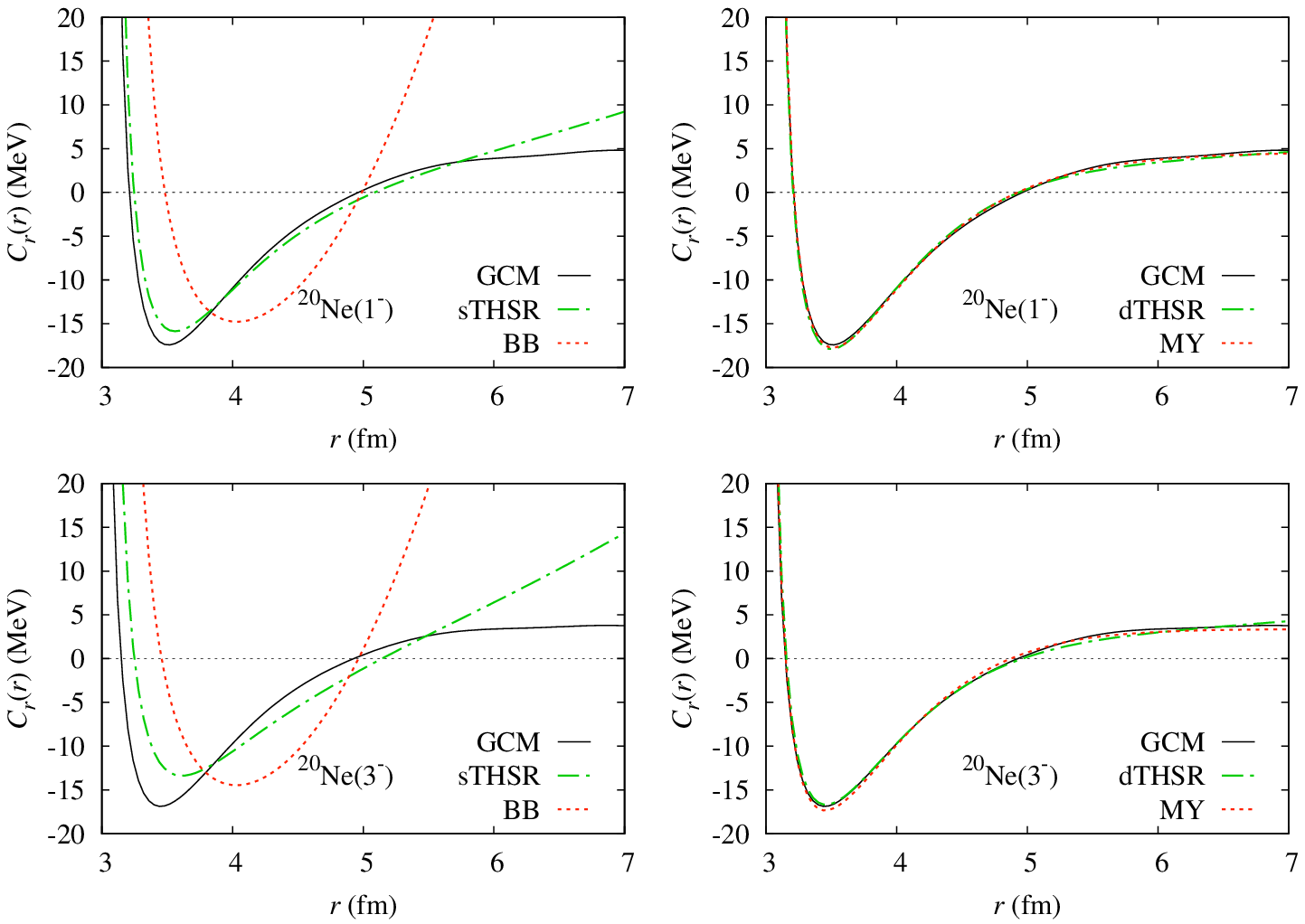} 	
\end{center}
\vspace{0.5cm}
  \caption{\label{fig:ne20-VE-np}
Same as Fig.~\ref{fig:ne20-VE} but for the $K=0^-_1$ band states of $^{20}$Ne.
}
\end{figure}

\begin{figure}[tb]
\begin{center}	
	\includegraphics[width=13.5cm]{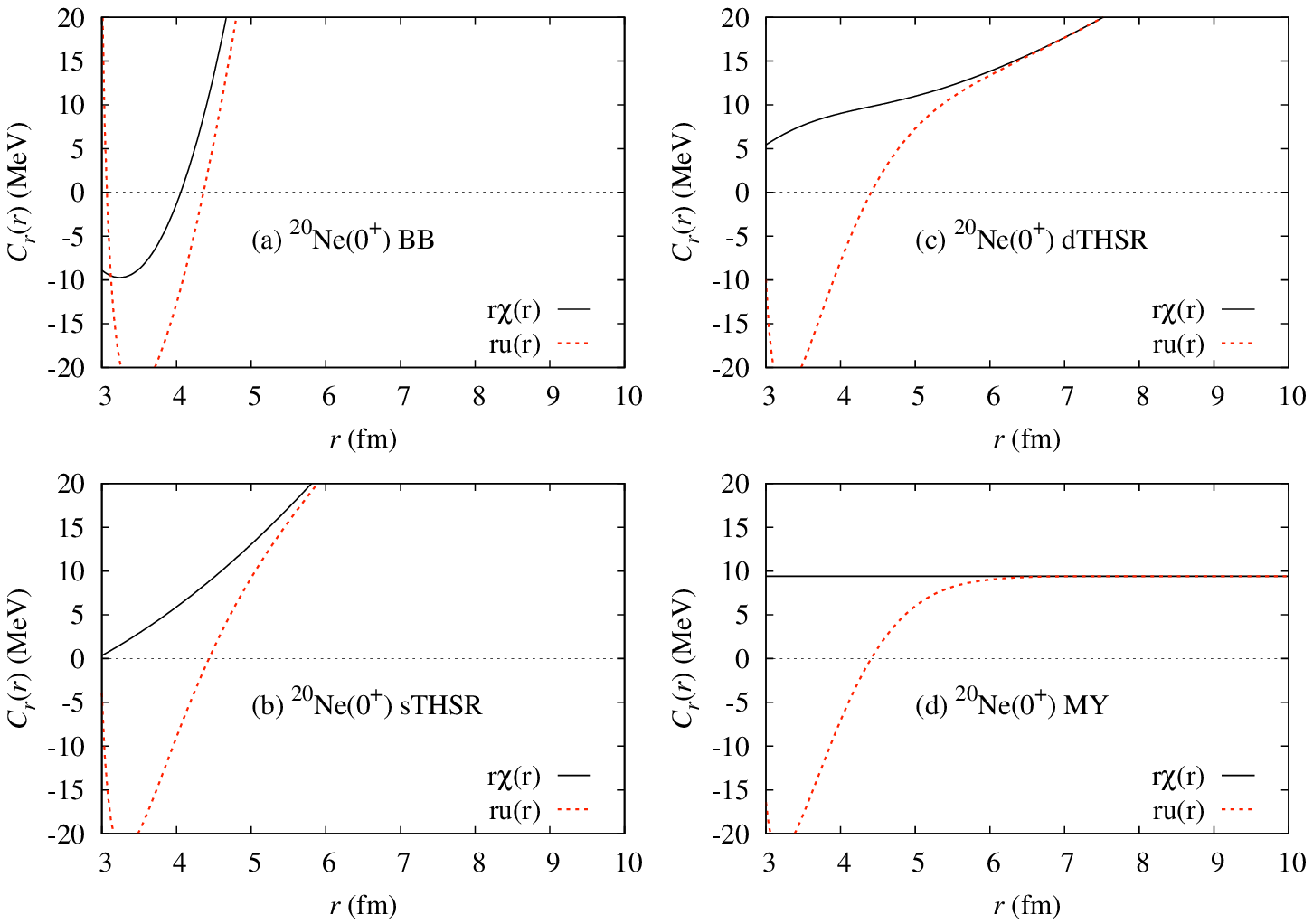} 	
\end{center}
\vspace{0.5cm}
  \caption{\label{fig:ne20-VE-chi}
Radial curvature ${\cal C}_r(r)$ of $r\chi_l(r)$ of the optimized trial 
functions for $^{20}$Ne($0^+_1$)\.
}
\end{figure}

\begin{figure}[tb]
\begin{center}	
	\includegraphics[width=13.5cm]{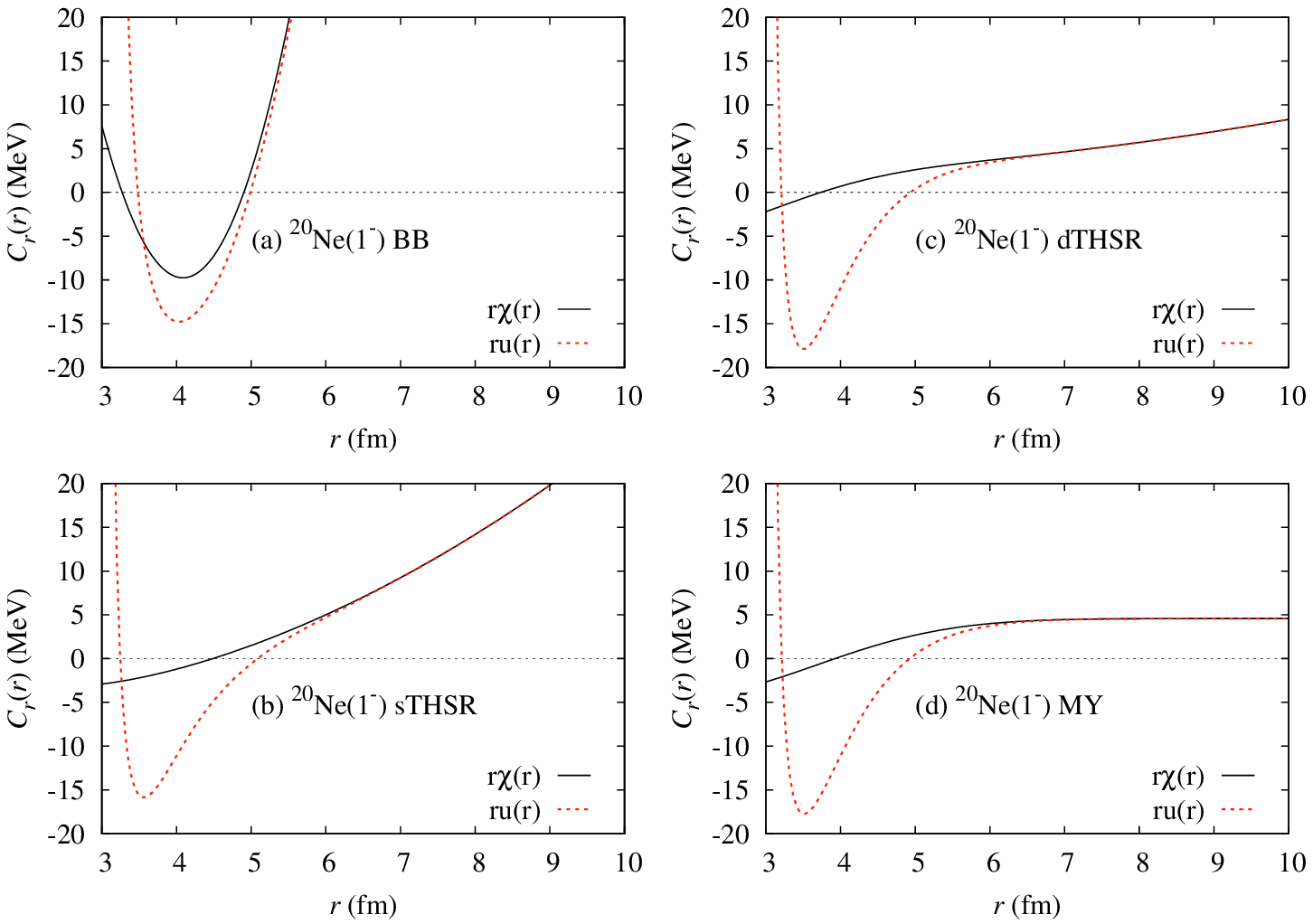} 	
\end{center}
\vspace{0.5cm}
  \caption{\label{fig:ne20-VE-chi-l1}
Radial curvature ${\cal C}_r(r)$ of $r\chi_l(r)$ of the optimized trial 
functions for $^{20}$Ne($1^-_1$)\.
}
\end{figure}

To discuss the damping behavior of the outer tail more quantitatively, 
we analyze the radial curvature ${\cal C}_r(r)$ of $ru_l(r)$ and $r\chi_l(r)$ defined 
in Eq.~\ref{eq:C}. 
In Figs.~\ref{fig:ne20-VE} and \ref{fig:ne20-VE-np}, the $r$ dependence of ${\cal C}_r(r)$ of $ru_l(r)$ for the 
optimized trial functions is compared with that for the GCM wave function. 
${\cal C}_r(r)$ of $r\chi_l(r)$ for the trial functions of $^{20}$Ne($0^+_1$) and  $^{20}$Ne($1^-_1$) 
is also shown in Figs.~\ref{fig:ne20-VE-chi} and \ref{fig:ne20-VE-chi-l1}.

${\cal C}_r(r)$ of $ru^{\rm GCM}_l(r)$ for the GCM wave function 
in the tail region shows rather flat behavior 
with a finite offset.
For $^{20}$Ne($0^+_1$), ${\cal C}_r(r)=5\sim$10 MeV in the outer tail region
and it is much larger than ${\cal C}_r(r)$ of $^8$B($0^+_1$) because of the larger $\alpha$ separation energy ($-E_r=5.92$ MeV) and 
the larger Coulomb barrier in $^{20}$Ne($0^+_1$). Also for $^{20}$Ne($4^+_1$) and $^{20}$Ne($8^+_1$), 
${\cal C}_r(r)$ in the tail region is as large as $C_r=5\sim$10 MeV 
because of the Coulomb and centrifugal barriers.
It is obvious that the BB and sTHSR wave functions fail to describe such the behavior of $C_r(r)$. 
The dTHSR gives the better result but it is difficult to fit the plateau of $C_r(r)$ in the outer tail region except for the case that ${\cal C}_r(r)$ is small enough.

As for $^{20}$Ne($1^-_1$) and $^{20}$Ne($3^-_1$) in the $K^\pi=0^-_1$ band, 
${\cal C}_r(r)$ of $ru^{\rm GCM}_l(r)$ of the GCM wave function in the tail region
is relatively smaller than the $K^\pi=0^+_1$ band states as ${\cal C}_r(r)<5$ fm because of the
small separation energy and the rather low centrifugal barrier. Moreover, since 
the surface peak is shifted outward because of the stronger antisymmetrization effect in the $K^\pi=0^-_1$ band states, 
the position crossing the ${\cal C}_r(r)=0$ line shifts to the larger $r$ region compared with the $K^\pi=0^+_1$ band states. 
In such the case, the sTHSR and dTHSR wave functions can give rather gentle slope of ${\cal C}_r(r)$, and therefore, 
it is easier to describe the $K^\pi=0^-_1$ band states than the $K^\pi=0^+_1$ band states.
However, ${\cal C}_r(r)$ of the dTHSR is a gradually increasing function in the outer region and it gives a slightly 
steeper slope of ${\cal C}_r(r)$ than that of the exact solution, 
in particular, of $^{20}$Ne($3^-_1$). 
On the other hand, the YT function can fit the flat region of ${\cal C}_r(r)$ in the outer region,
and therefore it gives better results in reproducing 
$u_l(r)$ and ${\cal C}_r(r)$ even for the $K^\pi=0^-_1$ band states than the dTHSR as well as for the 
$K^\pi=0^+_1$ band states. This indicates that the Yukawa-like tail is essential in 
the $\alpha$-cluster states of $^{20}$Ne. 

The radial curvature ${\cal C}_r(r)$ can be regarded as the effective potential with the offset $-E_r$ as  
${\cal C}_r(r)=V^{\rm eff}(r)-E_r$ in which the $\alpha$ cluster is confined and moving in the 
relative wave function. 
As seen in Figs.~\ref{fig:ne20-VE-chi} and \ref{fig:ne20-VE-chi-l1}, the effective potential for 
the physical wave function $ru_l(r)$ at the surface region
is dominantly described by the antisymmetrization effect, and it is quite different 
from that for the original
relative wave function $r\chi_l(r)$ before the antisymmetrization. 
On the other hand, in the outer region, 
the effective potential for $ru_l(r)$ is consistent with that for $r\chi_l(r)$.
As clearly seen, the effective potential shown by ${\cal C}_r(r)$ for the exact solution 
$ru^{\rm GCM}_l(r)$ can not be described by the Harmonic oscillator potential, 
but it shows a rather normal shape of the Coulomb and centrifugal barriers 
with a certain finite range attraction. 
Therefore, it is not surprising that the YT function with the Yukawa tail gives the best fit 
rather than the sTHSR with the Gaussian tail.
As explained before, the dTHSR wave function is not necessarily successful to 
describe the Yukawa-like tail except for the small ${\cal C}_r(r)$ case. 

These results indicate the following facts. The fixed-range Gaussian tail in the BB wave function is not suitable to reproduce the outer tail of cluster wave functions in $^{20}$Ne system.
The Gaussian tail with the adjustable range is better than the fixed-range Gaussian as argued 
in Refs.~\cite{Zhou:2012zz,Zhou:2013ala}.
However, the Yukawa-type tail can describe the outer tail well 
rather than Gaussian tail. The dTHSR wave function does not work so well as the YT function except for $^{20}$Ne($1^-_1$), because the cluster states of $^{20}$Ne
are not weakly bound states but they are rather well "bound" states 
because of the larger separation energy and/or 
higher centrifugal and Coulomb barriers than $^8$Be($0^+_1$).

\section{Discussion and Summary} \label{sec:summary}
We analyze the $\alpha$-cluster wave functions in the cluster states of $^8$Be and $^{20}$Ne 
by comparing various types of trial functions such as 
the BB, sTHSR, dTHSR, and YT functions
with the exact wave function obtained by the GCM calculation.
The relative wave functions in the BB, sTHSR, dTHSR, and YT functions are
given by the localized Gaussian with the fixed range, the spherical Gaussian, 
the deformed Gaussian, and the Yukawa-tail function, respectively. 
By investigating the squared overlap of the trial functions with the GCM wave function, 
we study how well the trial functions can describe the 
exact cluster wave function.
Compared with the SU(3) shell-model limit wave function, the description of the suppressed inner part and the 
enhanced surface peak of the physical relative wave function $ru_l(r)$ 
is improved with the BB wave function. The better result is obtained 
with the sTHSR wave function than the BB wave function, and further improvement is given with 
the dTHSR wave function because these wave functions can describe better the outer tail part. 
The YT function gives almost equal quality to the dTHSR wave function for 
$^8$Be($0^+_1$) and $^{20}$Ne($1^-$), and even better description for such states 
as $^{20}$Ne($0^+$), $^{20}$($2^+_1$), and  $^{20}$($3^-_1$).
This result indicates that the outer tail of $\alpha$ cluster states
is characterized by the Yukawa-like tail rather than the Gaussian tail.

The relative wave functions in the $\alpha$-cluster states of $^8$Be and $^{20}$Ne 
are characterized by three parts, the oscillating inner part, the enhanced surface peak, and 
the outer tail. 
In the $\alpha$-cluster states, the inner part is suppressed while the surface peak
is relatively enhanced because of the antisymmetrization effect between clusters. 
The nodal structure in the inner region and 
the enhanced peak structure at the surface in the physical wave function 
$ru_l(r)$ are not so sensitive to the original trial functions $r\chi_l(r)$ before the 
antisymmetrization because of the strong antisymmetrization effect between clusters. 
The outer tail is caused by the quantum penetration and its asymptotic behavior 
is well defined. In the weakly bound $\alpha$ cluster states with the small 
$\alpha$ separation energy and the low angular momentum $l$, the wave function 
is slowly damping in the outer region and it has the remarkably long outer tail.

To get a good approximation of the exact solution (GCM wave function) for 
the weakly bound $\alpha$-cluster states,  
it is essential to fit the outer tail part, in particular, its slow damping behavior.
On the other hand, 
the inner and peak parts are relatively less important in the fitting because they are
 mainly determined by the antisymmetrization effect.
The outer tail part is the slowly damping function which characterizes the
almost zero-energy free $\alpha$ gas behavior.
This "free $\alpha$-cluster gas part" in the outer region
is understood as the delocalization of cluster. 
It should be pointed out that the origin of this delocalization, i.e.,  
the "free $\alpha$-cluster gas part" in the outer region, 
is the quantum penetration and it is the obvious consequence of the weakly bound system. 
In the asymptotic region where the nucleus-nucleus interaction vanishes, 
the damping of the outer long tail in the free gas region is well defined by the 
$\alpha$ separation energy. In the ideal case that the separation energy is
small and the Coulomb and centrifugal barriers are not high, the
dTHSR wave function can describe the exact wave function
fairly well because it has the suitable form 
to fit the long tail in the outer free gas region. However, 
the success of the dTHSR is the mathematical result of the 
fact that the dTHSR can describe the Yukawa-like tail 
rather than the Gaussian tail.

Compared with the outer tail region free from the antisymmetrization, 
it is difficult to discuss physical meaning of the inner region and even of the 
surface peak region, because the physical wave function in these regions is not 
so sensitive to the original trial function before the antisymmetrization 
but it is strongly affected 
by the antisymmetrization effect. 
Instead, in the outer region, 
the alpha cluster wave function is well defined, and the features of 
the original model wave functions before the antisymmetrization 
are reflected more directly in the physical wave function after the antisymmetrization.
The physical meaning of the delocalization of clusters is clearly given in this region 
by the outer long tail caused by the quantum penetration, which can be regarded as 
the almost zero-energy "free $\alpha$ gas".

It should be stressed that 
the dTHSR with the large deformation has the 
damping behavior of the tail part quite different from the Gaussian tail of the sTHSR. 
%In the case of an enough large deformation, the dTHSR gives the slowly damping tail similar to 
%Yukawa-like tail.
%The radial curvature $C_r(r)$ can be regarded as the effective potential with the offset $-E_r$ 
%$C_r(r)=V^{\rm eff}(r)-E_r$ in which the $\alpha$ cluster is confined and moving in the 
%relative wave function. 
The effective potential evaluated by the radial curvature $C_r(r)$ 
for the physical relative wave function of the exact solution does not show the feature of 
the Harmonic oscillator potential, but it shows a normal shape of 
the Coulomb and centrifugal barriers with a certain finite range attractive potential.
Therefore, it is not surprising that the YT function with the Yukawa tail gives the best fit 
among the present trial functions rather than the sTHSR having the Gaussian form. 
The dTHSR wave function can successful describe the Yukawa-like tail in the 
exact solution in the case that $C_r(r)$ is small in the outer tail region.
If we apply the "container picture" proposed in Ref.~\cite{Zhou:2013eca}, it is better to consider 
a constant barrier with a small hight rather than Harmonic oscillator potential as the 
confining potential to understand the slowly damping Yukawa-like tail of the cluster wave function in the physical region. 

In Refs.~\cite{Zhou:2013eca}, 
the container picture has been proposed to understand nuclear clustering such as the $3\alpha$ state
in $^{12}$C system. 
It has been shown that the RGM wave function for $^{12}$C($0^+_2$), which corresponds to the 
exact solution of the $0^+_2$ state within the $3\alpha$ cluster model space, can be described 
fairly well by a single deformed THSR wave function \cite{Zhou:2013eca,Funaki:2009zz}. The result of the deformed 
THSR is much better than the spherical THSR function. It should be noted that 
the deformation of the optimized THSR wave function for $^{12}$C($0^+_2$) is large
with about 1:3 ratio. 
One should be careful again that the tail behavior of the 
$J^\pi$-projected wave function of the largely deformed Gaussian is different from the Gaussian tail. 
If we omit the angular momentum coupling and consider only 
the $^8$Be($0^+$)+$\alpha$ component, the radial wave function of the $\alpha$ cluster in the outer tail 
region is given by the form similar to the $2\alpha$ system discussed in the present work.
The largely deformed THSR wave function may not give the Gaussian tail but it might 
describe the slow damping behavior of the outer tail, in which the $\alpha$ clusters 
behave as almost zero-energy "free $\alpha$ gas".
In this outer tail region, the dynamics is governed by the quantum penetration 
though the small Coulomb barrier free from potential, and it is determined dominantly by the 
$\alpha$ separation energy.
One should also take care about the "physical region" in the $3\alpha$ system. Since the inner part 
is strongly affected by the antisymmetrization between clusters and also by 
the orthogonality to the lowest state $^{12}$C($0^+_1$), 
it is not easy to clearly mention the physical meaning of the inner part. 
Instead, the delocalization of clusters can be defined in the outer region free from the effects of the antisymmetrization and orthogonalization. 
As the separation energy becomes small, the outer tail part of the wave function 
becomes more and more important. In such a case, 
the delocalization could be 
characterized by the slowly damping long tail of the almost zero-energy "free $\alpha$ gas" as mentioned before.
It may be useful to investigate the details of the outer tail of the $3\alpha$ state, 
in particular, its damping behavior
in order to understand the proper shape of the confining effective potential in the container picture. 

In general, 
the delocalization occurs in weakly bound cluster states
definitely at least in the outer region. Even though the Gaussian tail of the 
single sTHSR function is not sufficient to 
perfectly reproduce the details of the outer tail, in particular, the damping behavior, 
it is usually better in description of the long tail than the
fixed-range Gaussian of the single BB wave function.
Further drastic improvement in the description of the outer tail can be obtained by 
the dTHSR function.
It should be also pointed that the dTHSR wave function is not only 
mathematically useful but also it is a powerful tool to investigate the 
$\alpha$ cluster states because the total microscopic wave function of the system 
is given in the quite simply form and it is easy to practically calculate
the energy expectation value of the total wave function.
In contrast to the dTHSR function, 
the present analysis of the YT relative wave function is just a mathematical game, 
and it is not practical to use the YT function in the actual microscopic calculation 
of many-body systems.

\section*{Acknowledgments} 
The authors would like to thank Dr.~Funaki, Dr.~Zhou, and Dr.~Suhara
for fruitful discussions.
The computational calculations of this work were performed by using the
supercomputers at YITP.
This work was supported by 
JSPS KAKENHI Grant Numbers 22540275 and 26400270.

\appendix
\section{$\chi^{\rm dG}_l(r)$ in large deformation limit} \label{sec:large-deformation}
In the case of largely deformed Gaussian 
with $\sigma_z \gg \sigma_\perp$, the curvature ${\cal C}_r(r)$ 
of $r\chi^{\rm dG}_l(r)$
can become small 
in the outer region as explained below. $\chi^{\rm dG}_l(r)$ in Eq.~\ref{eq:dGl} for even $l$ is rewritten as
\begin{eqnarray}
\chi^{\rm dG}_l(r)&\propto &
 \int^\pi_0 P_l(\cos \theta)\exp \left( -\frac{r^2}{\sigma^2_\perp}+\frac{r^2}{\Delta}\cos \theta^2 \right)
 \sin \theta d\theta \nonumber \\
 & = & \exp \left( -\frac{r^2}{\sigma^2_z}\right) 
 \int^\pi_0 P_l(\cos \theta) \exp \left( -\frac{r^2}{\Delta}\sin \theta^2 \right)
 \sin \theta d\theta.
\end{eqnarray}
In the $\sigma_z \gg \sigma_\perp$ case, in the asymptotic region of 
$r \gg \sigma_\perp$, 
only small $\theta$ region contributes to the integral, and then 
$\chi^{\rm dG}_l(r)$ can be approximately estimated as 
\begin{eqnarray}
\chi^{\rm dG}_l(r)&\approx& 
\exp \left( -\frac{r^2}{\sigma^2_z}\right) 
2\int^\infty_0 P_l(1) \exp \left( -\frac{r^2}{\Delta}\theta^2 \right)
\theta d\theta \nonumber\\
&=& \frac{\Delta}{r^2}\exp \left( -\frac{r^2}{\sigma^2_z}\right). 
\end{eqnarray}
In the $r\lesssim\sigma_z$ or $r\approx \sigma_z$ region, 
the $\exp \left( -\frac{r^2}{\sigma^2_z}\right)$ term changes gradually
and $\chi^{\rm dG}_l(r)$ approximately has the $\frac{1}{r^2}$ behavior,
and therefore, the curvature ${\cal C}_r(r)$ of $r\chi^{\rm dG}_l(r)$ is 
roughly estimated as 
${\cal C}_r(r)\approx \frac{\hbar^2}{\mu r^2}$. 
Namely, for $r\chi^{\rm dG}_l(r)$ with 
$\sigma_z \gg \sigma_\perp$, the curvature ${\cal C}_r(r)$ goes to 0 
in the $\sigma_\perp \ll r \lesssim \sigma_z$ region. 
It means that the deformed Gaussian should be a better trial function that can 
efficiently describe the slow damping behavior of the outer tail than the spherical Gaussian having 
more rapid damping of the Gaussian tail. 

The $\frac{1}{r^2}$ behavior in the $r\lesssim\sigma_z$ region of $\chi^{\rm dG}_l(r)$ with $\sigma_z\gg \sigma_\perp$ 
can be more intuitively understood by a cylinder
picture. The angle average of a cylinder with a diameter $\sigma_\perp$ at $r\gg \sigma_\perp$ is 
approximated as $\pi\sigma^2_\perp/4\pi r^2$.

\end{document}